\documentclass[a4paper,fleqn]{cas-dc}
\usepackage[numbers]{natbib}

\usepackage{xcolor}
\usepackage{amssymb}
\usepackage{hyperref}
\usepackage{xr-hyper}
\usepackage{siunitx} 
\usepackage{float}

\usepackage{ulem}

\def\tsc#1{\csdef{#1}{\textsc{\lowercase{#1}}\xspace}}
\tsc{WGM}
\tsc{QE}

\begin{document}
\let\WriteBookmarks\relax

\newcommand{\silvana}[1]{\textcolor{green}{#1}}
\newcommand{\sander}[1]{\textcolor{magenta}{#1}}
\newcommand{\jer}[1]{\textcolor{blue}{#1}}

\externaldocument{SI_Section}

\shorttitle{Emergent non-reciprocal pair interactions in catalytically propelled active polyhedra}

\shortauthors{}  

\title [mode = title]{Emergent non-reciprocal pair interactions in catalytically propelled active polyhedra}  


\tnotemark[1]


\tnotetext[1]{\textit{$\dagger$ Electronic supplementary information (ESI) available. See DOI:}} 

\author[1]{Silvana A. Caipa Cure}
\ead{caipa@physics.leidenuniv.nl}
\credit{Conceptualization, data curation, methodology, formal analysis, investigation, writing – original draft, visualization}

\author[2]{Sander C. Kammeraat}
\ead{kammeraat@lorentz.leidenuniv.nl}
\credit{Conceptualization, Data curation, methodology, formal analysis visualization, investigation, writing – original draft}

\author[3]{Jérôme Hem}
\ead{jerome.hem@univ-lorraine.fr}
\credit{Methodology, Data curation, investigation,  formal analysis visualization, writing – review and editing}

\author[2]{Silke Henkes}
\ead{shenkes@lorentz.leidenuniv.nl}
\credit{Conceptualization, methodology, formal analysis, writing – review and editing}

\author[1]{Daniela J. Kraft} [orcid=0000-0002-2221-6473]
\ead{kraft@physics.leidenuniv.nl}
\credit{Conceptualization, methodology, formal analysis, resources, writing – original draft and editing, supervision, funding acquisition}
\cormark[1]
\cortext[1]{Corresponding Author}

\affiliation[1]{organization={Huygens Laboratory}, addressline={Niels Bohrweg 2}, city={Leiden},country={The Netherlands}}

\affiliation[2]{organization={Lorentz Institute}, addressline={Einsteinweg 55}, city={Leiden},country={The Netherlands}}

\affiliation[3]{organization={Institut Jean Lamour, CNRS, Université de Lorraine}, addressline={54011}, city={Nancy},country={France}}

\begin{abstract}
Particle shape plays a fundamental role in the dynamics and pair interactions of active particles. Yet, experimental measurements of anisotropic active particles are largely lacking. Here, we introduce an experimental system that allows us to observe and measure the effective pair interactions of self-propelled active colloidal polyhedra. We realize such a system by 3D micro-printing quasi-2D polyhedral colloids with edge number $N$= 3,4,5,6 and $\infty$ and induce propulsion by partially coating them with an iridium catalytic cap and dispersing them in hydrazine. Despite their different geometries, all shapes exhibit similar single-particle positional dynamics. Stochastic Force Inference reveals predominantly radial, non-reciprocal pair interactions with similar attractive and repulsive regimes for all particle shapes at center-to center distances of about two particle diameters. 
A minimal plausible phoretic active-Brownian-particle model with a disc shape reproduces the experimentally observed behavior and identifies the physical origins: concentration-dependent slowing down sets their dominant angular symmetry, while the spatial geometry of the chemical coupling and phoretic attraction recover their relative magnitudes and angular dependence. 
We find that the shape anisotropy present in experiments only modulates the magnitudes and angular dependence of these interactions, and that these variations become more pronounced the stronger the anisotropy.
Our results demonstrate the interplay of chemically mediated interactions and particle geometry-induced effects at intermediate interaction distances. 

\end{abstract}

\begin{graphicalabstract}
\includegraphics{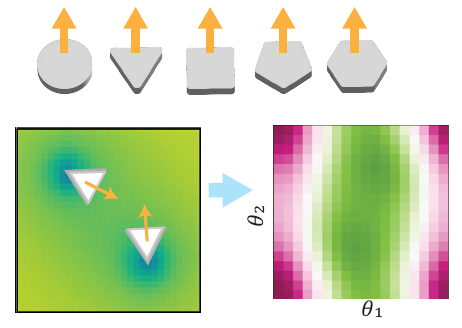}
\end{graphicalabstract}

\begin{highlights}
\item {Colloidal polyhedra are fabricated via 3D printing and functionalized to display catalytic self propulsion at ultra-low levels of hydrazine fuel.}
\item {Stochastic Force Inference is used to infer effective pair-wise forces among chemically active polyhedra yielding non-reciprocal interactions.}
\item {A phoretic active Brownian particle model that includes the concentration field dynamics and the concentration-dependent active velocity is used for mapping the effective inferred forces and torques to underlying mechanisms.}
\item {Long-range chemical fields, extending over distances of up to five particle diameters, produce similar attractive and repulsive interaction patterns across active colloidal particles of different shapes.}
\item {Geometrical anisotropy modulates interaction magnitudes and introduces additional aligning and anti-aligning torques.}
\end{highlights}

%\nocite{*}

% Keywords

\maketitle

% Main text

\section{Introduction}
Microscopic synthetic active particles display a wide range of collective dynamics, including non-equilibrium clustering~\cite{Theurkauff2012, palacciLivingCrystalsLightactivated2013, Weitz2015}, swarming~\cite{Yan2016, Shelke2026}, chain formations~\cite{Yan2016} and quorum sensing~\cite{Lefranc2025}. These emergent dynamics arise from the interplay of short- and long-range interactions between the particles. Short-range interactions, which are also present in passive colloids and include van der Waals and electrostatic interactions, act only over distances of a few nanometers from the particle surface. For active particles,  the propulsion can generate additional long-range interactions which can include hydrodynamic, phoretic and substrate-mediated osmotic interactions which can extend over several particle diameters~\cite{Liebchen2021}. For example, experimental measurements on Janus spheres powered by surface catalytic reactions, show that self-generated chemical gradients mediate attractive phoretic interactions between the particles promoting the formation of dynamic clusters~\cite{Theurkauff2012}.

Experimental measurements of the interactions in active colloidal systems remain scarce and have focused primarily on spherical active colloids. Existing approaches include using passive tracers to distinguish between phoretic and osmotic contributions to the particle interactions~\cite{Hardikar2024,Boniface2024}, analyzing particle trajectories to extract interaction-induced changes in translational speed and orientation~\cite{Sharan2023,Singh2024}, and, more recently, inferring effective pair interactions from large ensembles of self-propelled Janus spheres using Stochastic Force Inference (SFI)~\cite{Hem2025}.

Much less is known about how the particle geometry modifies the interactions of active particles, despite its fundamental role in the collective behavior of active matter~\cite{Wensink2013,Damasceno2012,Fily2014,vanDamme2019,Moran2022, Riedel2024, Riedel2025}. 
Already in the absence of hydrodynamic or chemical interactions, particle geometry can strongly modify active-particle encounters and phase behavior. For example, simulations of a system of active 2D polygons interacting through a purely repulsive Weeks–Chandler–Andersen (WCA) potential have shown that the shape and the orientation of the propulsion force relative to the particle body strongly affect the collision dynamics. As a result, the critical densities for the onset of motility-induced phase separation can vary substantially between shapes and be higher or lower than those for isotropic colloids~\cite{Moran2022}. 

Previous experimental studies have demonstrated shape-induced locking between active colloidal crescents~\cite{Riedel2024}, hydrodynamic interactions of pairs of AC-field driven discs~\cite{Katuri2022}, and aspect-ratio-dependent pusher-puller behavior in active colloidal rods~\cite{Shelke2026}. While these isolated examples establish that geometry can influence active-particle interactions, they also showed that interactions often extend beyond steric effects only. Even more so, since fabricating monodisperse samples of self-propelled particles with controlled geometries is challenging, the effect of shape anisotropy on the pair interactions has not been systematically investigated across a controlled series of particle symmetries.

Here, we introduce an experimental system of active colloids fabricated by two-photon-based 3D microprinting to address this gap. We experimentally measure and quantify the pair interactions of active polygonal colloids with $N$= 3, 4, 5, 6 and $\infty$ edges and induce their 2D propulsion through the catalytic decomposition of hydrazine at a thin iridium cap. 
We characterize their single particle dynamics and extract effective pair interactions using Stochastic Force Inference, for a medium interparticle distance $r\approx 2\sigma$ with $\sigma$ the particle diameter. Despite there only being minor differences between the different polygons, we find that the interactions are predominantly radial and strongly depend on the relative orientation of the two particles in a non-reciprocal manner. 
To investigate their physical origin, we construct a minimal plausible phoretic active Brownian particle (ABP) model that explicitly accounts for the evolving fuel concentration field and the concentration dependence of the propulsion speed. Applying the same SFI framework to the simulated trajectories reproduces the experimentally observed interaction-induced radial and tangential velocities, suggesting that concentration-dependent slowing down determines their common angular symmetry, while the spatial geometry of the chemical coupling and phoretic attraction determine their relative magnitudes and finer structure. 
The experimentally inferred angular velocities for the squares and the triangles have a richer dependence on the particle orientations compared to those for discs and the pentagons. This suggests the presence of other contributions, such as shape anisotropy and hydrodynamic effects that can induce (anti)alignment of the polarities. 

\section{Experimental Materials and Methods}

\subsection{Materials}
Hydrazine (H\textsubscript{2}N\textsubscript{4}, $35\%$, in water), propylene glycol monomethylether acrylate (PGMEA, $>99.5\%$) and Novec 7100 Engineering Fluid (HFE-7100) were purchased from Sigma-Aldrich. Promosol DR3 was purchased from Inventec. 2-propanol (IPA, $9,99\%$) was purchased from VWR Chemicals. The Ir target (57mm dia. x 0.3mm thick, 99.8+\%) was purchased from Testbourne B.V. All materials were used as received unless stated otherwise. All solutions were prepared from deionized water with 18.2M$\Omega$ cm resistivity, using a Millipore Filtration System (Milli-Q Gradient A10).

\subsection{Colloidal fabrication}

\subsubsection{Particle fabrication} 
Colloidal polygons were fabricated using a commercial two-photon polymerization (2PP) 3D micro-printer (Photonic Professional Gt, Nanoscribe GmbH) equipped with a 63$\times$ oil-immersion objective (Zeiss, NA=1.4) in dip mode and the commercially available photoresist IP-Dip. 
The polygonal shape with the desired number of edges $N$ was designed such that it was enclosed in a circle with diameter $d$=8-10 \si{\micro \meter} and the thickness $w$ was set to yield an aspect ratio of $\alpha=w/d=0.2-0.25$ (see Fig.\ref{fgr:fig1}A (i)). The polygons were designed using Autodesk Inventor and rendered for printing in Describe. The structures were printed standing up on one of their edges (see Fig.\ref{fgr:fig1}A (ii)) onto a fused silica substrate functionalized with 3-(Trimethoxysilyl)propyl methacrylate (TPM) using a modified version of the method in Ref.~\cite{vanDerWel2017}, to improve the adhesion of the features and avoid them from falling. 
During printing, a slicing distance of 0.2, contour count of 1 and hatching distance of 0.2 were used. The laser power was set to 0.6\% of the full power.
After printing, the particles were developed by 30 min submersion in PGMEA followed by a 2 min rinse in HFE-7100 or Promosol DR3 (depending on availability), which removes residual developer and prevents particles from detaching during drying.

\subsubsection{Implementation of self-propulsion}
The colloidal polygons were sputter-coated with a thin metallic layer of Iridium to integrate the propulsion mechanism. For this the particle's surface was first functionalized with a silicon oxide layer to enhance the reactivity of the catalytic layer~\cite{Gao_2014,Jang2012}. We used a high vacuum Leybold-Heraeus Z400 sputter-coater with custom modifications to obtain a 20 nm SiO\textsubscript{2} coating of all the exposed surfaces. In this system the target is close to the substrate and sputtering is performed at $\approx 10^{-5}$ mbar (Ar @ 54 sccm, Direct Current Electrode Positive (DCP): 1kV) yielding a robust layer that covers most of the particle. Next, we employed a low vacuum ($10^{-2}$ mbar) Cressington 208HR sputter-coater to  create a thin (20 nm) metallic cap of Ir on the top of the particles (see Fig.~\ref{fgr:fig1}A). 
Sputter-coated particles were recovered from the substrate via sonication and dispersed in water. Fig.~\ref{fgr:fig1}B shows SEM micrographs of the recovered shapes where the red dotted line indicates the edge that was attached to the substrate during printing and sputtering and hence was not coated with Iridium.

\subsection{Particle observation}
To characterize the surface of the particles after sputter-coating, scanning Electron Microscopy (SEM) images were taken with a Thermo-Fisher Apreo SEM. For this, particles were deposited onto a silicon SEM substrate and allowed to dry in air before imaging.
\\
Information on the dynamics of the metal-coated particles was extracted from brightfield microscopy observations in a custom made microscope holder in the presence of 0.0005\%v/v hydrazine (N\textsubscript{2}H\textsubscript{4}) in water. 
The microscope holder consisted of two aluminum rings that can be assembled together with two circular cover slips to form a chamber that is 1.5 cm in diameter and 1 cm in height. 
The density difference between the particles and solvent together with activity-induced interactions with the substrate ensured quasi-two-dimensional motion of the particles~\cite{ketzetziDiffusionBasedHeightAnalysis2020}.
A 0.0005\% v/v N\textsubscript{2}H\textsubscript{4} dilution was freshly prepared from a 1\%v/v stock for every experiment, to avoid depletion of the fuel due to contact with the atmosphere. The 1\%v/v stock was kept under Nitrogen gas and at 4$^\circ\mathrm{C}$. Particles were imaged using a Nikon Eclipse Ti microscope with  $20\times$ dry (NA=0.5) and a $10\times$ dry (NA=0.45) objectives depending on the measurement. The frame rate and duration of the experiment varied per sample and is specified per case. 

\section{Tracking and Numerical Analysis}
\subsection{Tracking}
\subsubsection{Low density sample for single particle dynamics} 
To study the single dynamics of particles, the samples were prepared at a very low density of active polygons (less than 5 particles per field of view of 655 \si{\micro \meter} x 655 \si{\micro \meter}) which were recorded at $20\times$ magnification for 90 s with a frame rate of 5 or 10 fps. We extracted the active velocity $U$, the translational $D_t$ and rotational $D_r$ diffusion coefficients from this data. To achieve this, the position of the particles $\mathbf{r}(t)$ was detected in each frame using pixel intensity and area thresholding in order to extract the center of mass of the features. The centers of mass were linked using the Crocker-Grier algorithm implemented for Python as Trackpy~\cite{Trackpy}, resulting in the trajectories displayed in Fig.~\ref{fgr:fig1}C. A total of  15-30 independent trajectories were recorded per shape.

From the trajectories, the mean square displacements (MSD) were calculated and fitted to $\langle \| \mathbf{r}(t+\Delta t)- \mathbf{r}(t) \|^2\rangle= 4D_{t}\Delta t+ U^{2} \Delta t^{2}$, where $\Delta t$ is the lag time in a range of $\Delta t_{min}= 1s $ and $\Delta t_{max}= 5s $, which is smaller than the rotational diffusion timescale but large enough to be in the ballistic regime. $D_r$ was obtained by correlating the unitary displacement vector of the particle $\mathbf{e}_{\varphi}(t) = \mathbf{U}(t) /  \| \mathbf{U}(t) \|$ from the velocity $\mathbf{U}(t) = (\mathbf{r}(t+\delta t) - \mathbf{r}(t)) / \delta t$ with $\delta t=1s$.
The correlations were averaged over at least 15 independent trajectories per shape and fitted to the exponential decay relation $\langle \mathbf{e}_{\varphi}(t)\cdot\mathbf{e}_{\varphi}(t+\Delta t)\rangle = e^{-D_{r} \Delta t }$.

\subsubsection{Medium density experiments for pair-wise interactions} 
Samples for the inference analysis were prepared at a higher packing fraction of active polyhedra $\phi\approx1\%$, and observed for a longer time, 10 min at 10 fps, in order to satisfy the statistical requirements. A $10\times$ objective was used to capture many particles in a single field of view while maintaining a sufficient density at which only pair interactions are observed. See Supplementary movie S1. 

To detect the orientation great care was taken to capture the edges of the polygon as accurately as possible so as to extract the inherent polarity of the particle at all times ($\dagger$ESI Fig. S2). An enhanced binarized image was used to distinguish the relevant features from the background. The change in curvature along the contour of the features was employed to detect the polygonal vertices throughout all the frames. These vertices were then related to the position of the center of mass $\mathbf{r}(t)$ and to the particle's intrinsic polarity $\varphi(t)$. 
Hexagons were excluded from these measurements because of the poor resolution of their very short edges at the employed magnification, resulting in unreliable polarity identification and tracking.

\subsection{Stochastic Force Inference}

To infer the interactions, we used the stochastic force inference (SFI) algorithm of Ref.~\cite{Frishman2020} which is specifically relevant for our system as it follows Langevin dynamics. The main steps can be summarized as follows: first a functional basis is defined, on which the measured drift is projected (Stratonovitch estimator) along the particle trajectories. Then, the regularized coefficients of the basis are obtained by an ortho-normalization procedure. The method has already been applied successfully to describe the dynamics of an underdamped particle~\cite{Bruckner2020} and to measure the interactions between electrophoretic flocking colloids~\cite{Hem2025}. 

For our analysis, we assume that all particles of one polygonal shape are identical, and more importantly that
they influence one another only through instantaneous and
pair-wise interactions. Although the two last assumptions significantly constrain the range of possible interactions, specifically those of chemical origin, they continue to be approximately relevant in our system, as we will demonstrate below. The fast diffusion of the chemical fuel compared to the particle velocity leads almost instantly to the establishment of a steady state profile of concentration around the particle, whereas the low to medium particle density that we study, favors mainly two body collisions. However it is useful to keep in mind the “effective” nature of the inferred interactions that only approximate the realistic ones.

In addition, we assume that the translational diffusion can be neglected because its contribution is much smaller than that of the active force, which we will justify below, and that the rotation coefficient $D_r$ is constant, so that the dynamics of particle $i$ for its position $\mathbf{r}_i(t)$ and polarity $\varphi_i$ can be defined as:
\begin{equation}
  \dot{\mathbf{r}}_i(t) = U_{\text{SFI}}\,\mathbf{e}_{\varphi_i}(t)
+ \sum_{j \ne i} \mathbf{v}(\mathbf{r}_i - \mathbf{r}_j, \varphi_i, \varphi_j)
\label{eq: SFI displacement EOM}
\end{equation}
\begin{equation}
    \dot{\varphi}_i(t) = \sum_{j \ne i} \omega(\mathbf{r}_i - \mathbf{r}_j, \varphi_i, \varphi_j)
+ \sqrt{2D_r}\,\xi_i(t),
\label{eq: SFI angular EOM}
\end{equation}

where $U_{\text{SFI}}$ is the self propulsion speed, $\mathbf{e}_{\varphi_i}(t)$ the unit vector pointing in the direction $\varphi_i$, $\xi_i(t)$ a Gaussian white noise with unit variance, and $(\mathbf{r}_j,\varphi_j)$ the position and polarity of particle $j$. The pair interactions are encoded in the effective additional velocity $\mathbf{v}$ that stems from the interaction forces
\begin{equation}
  \mathbf{v}(r,\theta_1,\theta_2)=v_r(r,\theta_1,\theta_2)\mathbf{e}_r+v_\theta(r,\theta_1,\theta_2)\mathbf{e}_\theta,
\label{eq: SFI v}
\end{equation}

which can be decomposed into a radial $v_r$ and tangential $v_{\theta}$ component with respect to the bond vector $\mathbf{r_{ij}}=\mathbf{r}_i - \mathbf{r}_j$ between particles $i$ and $j$, and the interaction-induced angular velocity $\omega(r,\theta_1,\theta_2)$. Here the coordinates $r_{ij},\theta_1,\theta_2$ represent the relative positions between the two particles $i$ and $j$, with the distance $r=|\mathbf{r}_{ij}|$ and the two relative angles $\theta_1,\theta_2$ with respect to the bond vector $\mathbf{r}_{ij}$ (see Fig~\ref{fgr:fig2}C). As a consequence, the quantities to be inferred are the rotational diffusion $D_r$, the self propulsion $U_{\text{SFI}}$ and the three functions $v_r(r,\theta_1,\theta_2)$, $v_\theta(r,\theta_1,\theta_2)$ and $\omega(r,\theta_1,\theta_2)$.

To project the pair interactions, we chose the basis functions to be products of radial and angular functions~\cite{Hem2025}. The radial functions are smooth localized functions defined by $g_l(r) = (1/l!) (r/r_w)^{l} e^{-r/r_w}$ 
with a maximum at $lr_w$ and indexed by the integer $l$. The angular basis is of Fourier type composed of terms such as $\cos([n-|k|]\theta_1+k\theta_2)$ or $\sin([n-|k|]\theta_1+k\theta_2)$ indexed by the integers $n$ and $k$. The resolution is then controlled by the number of radial functions, and the angular order defined by $\max(n)$. For our study we used $r_w=8$ \si{\micro \meter} with $l$ spanning $\in$ ${0,1,2,3,4,5}$ in order to cover a cell length of 70 \si{\micro \meter} for experiments and 50 \si{\micro \meter} for simulations, in which pair interactions are calculated. For the angular basis, we found that a 2nd order basis was sufficient for all particle shapes.

For all the experimental measurements, the data points are always blurred by additional sources of noise, that in our case affect mostly the polarity detection. To reduce this effect, we increase the amplitude of the measured signal by imposing a larger time interval, $dt$, between consecutive frames. However, to accurately track the particle positions, $dt$ cannot be too large either. A trade-off must therefore be made between increasing $dt$ and maintaining accurate particle tracking. Taking $U\approx3$ \si{\micro\meter\per\second}, we reasonably do not expect a spatial resolution lower than $k\sigma$ with $k=1/10$~\cite{Hem2025}.  This gives us an optimal time interval of $dt=k\sigma/U\approx0.3$ \si{\second} for our system. We therefore sample the experimental data for each shape such that $dt=0.3$ \si{\second}.

In addition, it is important to mention that uncertainties are also correlated with the amount of data analyzed. An idea of the magnitude of the uncertainty is given by the radial pair correlation function $g_l(r)$, which shows the probability density of encountering a neighboring particle at a distance $r$. An additional source of information are the histograms of the independent pair-wise events recorded for a specific distance $r_0$ with respect to $\theta_1, \theta_2$, which can can be found in $\dagger$ESI Fig. S5).
As discussed in section 4.2, the radial distribution $g_l(r)$ for all experimental measurements is significantly depleted in the short range $r<2\sigma$. This poor resolution affects dramatically the quality of the inference, as evidenced by the rapid and nonphysical variation of the interactions for $r<2\sigma$ ($\dagger$ESI Fig. S6), so that the interactions at this short range are inaccessible.   

\subsection{Numerical simulations}
To complement the experiments, we developed a minimal plausible, shape-independent, phoretic model of the Janus particles. In general, one would need to consider both the chemical field of the fuel and the field of the products of the catalyzed reaction since they could both have phoretic contributions. However, given that the consumption of the fuel and the creation of the products occur at the same location and that the decomposition of hydrazine by Ir has an extremely high reaction rate~\cite{Matyshak2022}, we consider for simplicity a single chemical field $c$ with the convention that the dynamics of $c$ follow the dynamics of the hydrazine concentration field. Then, we consider two commonly studied~\cite{Saha_2014,Popescu_2018,Liebchen_2019,Liebchen2021,Fadda_2023,Tucci_2024} effects on the particle: the first is phoresis along the product concentration gradient, described by the surface-averaged slip velocity $-\beta \nabla c(\mathbf{r}_i,t)$, with $\beta$ the phoretic mobility. We corroborated this effect by fitting the typical $\propto r^{-2}$ scaling expected for phoretic attraction~\cite{Liebchen2021}, to tracer particles moving towards a fixed polygon, see Fig.~\ref{fgr:fig3}B. The second is alignment of the orientation of the particle opposite to the direction of the product concentration gradient (that is, aligning to the fuel gradient). This in turn is corroborated by the observation of particles in clusters being aligned outwards, see Fig.~\ref{fgr:fig3}A.  The alignment takes the form of a torque $J\left[ \mathbf{e}_{\varphi_i}(t) \times \nabla c\left(\mathbf{r}_i,t\right)\right]$, with $J$ the strength of the alignment.
Since the SFI  equations of motions (\ref{eq: SFI displacement EOM}) and (\ref{eq: SFI angular EOM}) assume an activity in the form of an Active Brownian Particle, where the polarity vector has its own dynamics, we model the phoretic effects on the particle in the following minimal way,
\begin{equation}
  \dot{\mathbf{r}}_i(t) = U(c)\,\mathbf{e}_{\varphi_i}(t)-\beta \nabla c(\mathbf{r}_i,t)-\sum_{j\neq i}Ae^{-|\mathbf{r}_i - \mathbf{r}_j|/a} \mathbf{e}_{ij}
\label{eq: phoretic displacement EOM}
\end{equation}
\begin{equation}
    \dot{\varphi}_i(t) = J\left[\mathbf{e}_{\varphi_i}(t) \times \nabla c(\mathbf{r}_i,t)\right]\cdot \mathbf{e}_z
+ \sqrt{2D_r}\,\xi_i(t).
\label{eq: phoretic angular EOM}
\end{equation}

This model is similar to the model by Liebchen et al.~\cite{Liebchen_2019}, but with three important differences.
First, the self-propulsion term in equation (\ref{eq: phoretic displacement EOM}) is not of constant magnitude, but depends on the concentration of the chemical field $c$. This is because the self-propulsion speed is known to sensitively depend on the concentration of fuel for the decomposition of hydrazine catalyzed by Iridium metal~\cite{Gao_2014}. To introduce this strong dependence of the speed on the concentration, we interpolate linearly between the  experimental measurements, see  Fig.~\ref{fgr:fig3}C and set the speed to zero outside the range. 

Second, modeling this speed dependence forces us to explicitly keep track of $c$ as a function of time and space, instead of integrating out its time dependence due to its fast dynamics~\cite{Liebchen2021}. In the experiments the reservoir consists of a large volume (600\si{\micro\liter}) of 0.0005\%v/v hydrazine that through diffusion constantly replenishes the observation plane with fresh fuel. The time evolution of $c$ is then governed by diffusion with diffusion constant $D_c$, depletion by the particles at a rate $k_d$ at their catalytic surface $S$ and a local renewal to the reservoir concentration $c_0$ at rate $k_r$: 
\begin{align}
\partial_t c(\mathbf{r},t) = &D_c\nabla^2c(\mathbf{r},t) + k_r\left[c_0-  c(\mathbf{r},t)\right]\nonumber \\ & -k_d\sum_{i} \frac{1}{S_i}\int_{S_i} ds\delta(\mathbf{r}- \mathbf{r}_i) 
    \label{eq: phoretic chemical field evolution}
\end{align}
Importantly, in the experiments and simulations $c_0$  remains on the left side of the peak in $U(c)$, such that $c(\boldsymbol{r},t)$ is in the regime where a decrease in concentration always leads to a decrease in speed (see Fig.~\ref{fgr:fig3}C).

Finally, since the particles are effectively attracting each other \textit{via} the phoresis term in equation (\ref{eq: phoretic displacement EOM}) due to the chemical gradients they create by locally depleting $c$, we need to include a repulsive interaction to prevent the particles from overlapping. In accordance with the experimental observation (see below) that the active colloids are repelling at a distance larger than their own size, we add a simple isotropic exponential repulsion with a strength $A$ and finite range $a$ in the direction opposite to the unit vector  $\mathbf{e}_{ij}$ pointing from particle $i$ to particle $j$. In Fig.~\ref{fgr:fig3}D we provide a graphical overview of the phoretic model.

We integrate equations (\ref{eq: phoretic displacement EOM}) and (\ref{eq: phoretic angular EOM}) using an Euler-Maruyma integration scheme. The field $c$ is discretized on a square grid and we use central finite differences to calculate the gradient and Laplacian of the field at each grid point, which are then used to evolve equation \ref{eq: phoretic chemical field evolution}. The grid points divide the simulation domain into bins, with the grid points as bin centers. The particles sense the concentration value of the bin they are in, that is, the value of the closest gridpoint. Specifically, since the chemical reaction takes place at the back of the particle, we resolve the interactions with the chemical field at the back side of the particle taking the average of the three bins (spanning $6$ \si{\micro\meter}) behind the particle, opposite to its polarity vector, see Fig.~\ref{fgr:fig4}B(i). This spatial sampling kernel retains the location and extent of the catalytic region relative to the particle polarity, even though the particles otherwise have no explicit shape.  The simulation domain has periodic boundary conditions. The complete simulation is implemented in the open-source simulation package JAMs~\cite{Kammeraat_JAMs}. As detailed below, parameters (see $\dagger$ESI Table 1) were matched using a loop of direct measurements and repeated SFI on the simulated data.

\section{Results and Discussion}

\subsection{Single particle dynamics of colloidal polygons}
\begin{figure*}[H]
\centering
  \includegraphics[width=.9\textwidth]{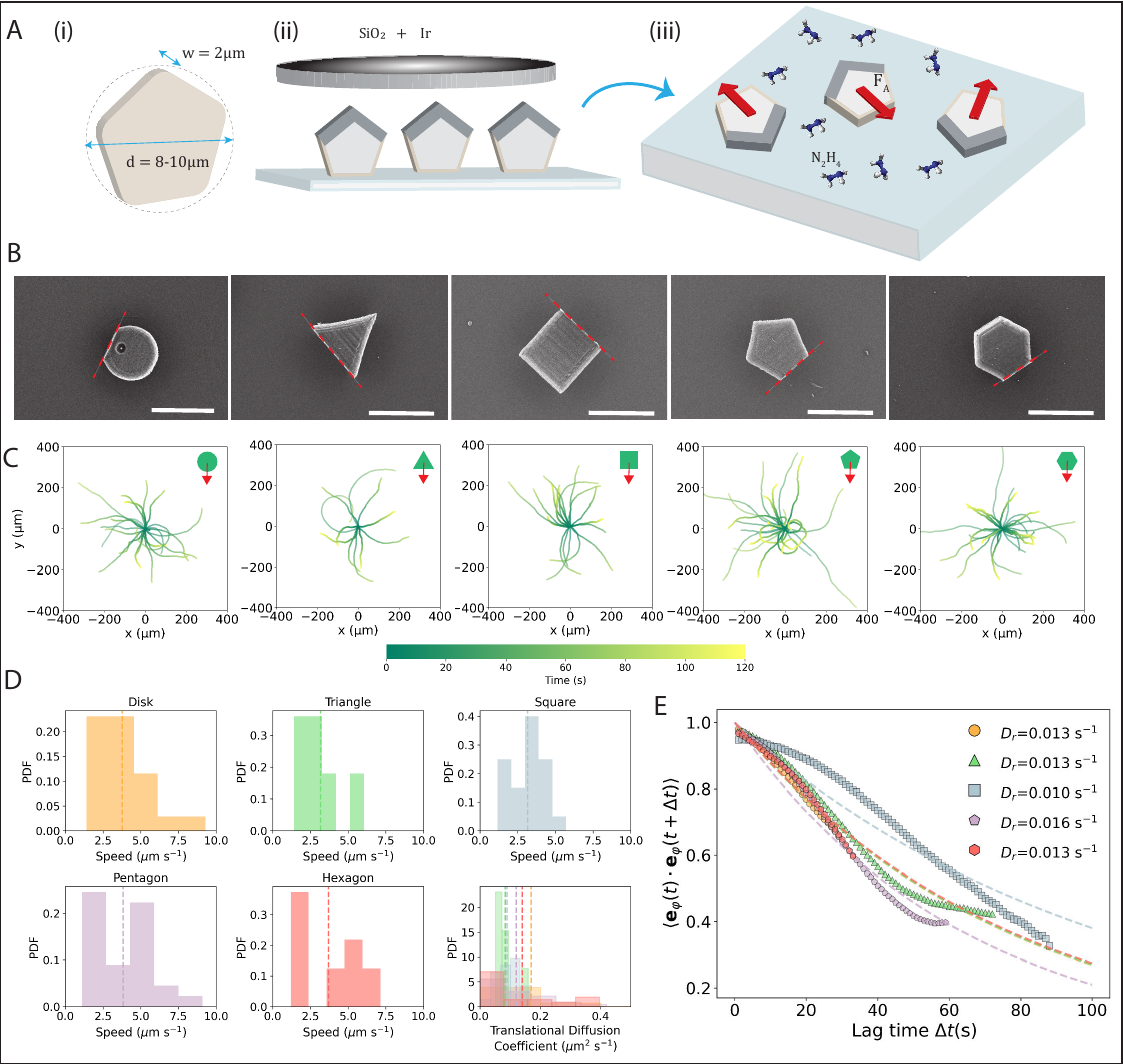}
  \caption{(A) Schematic of the design and fabrication process for 3D printed active polyhedra, consisting of (i) computer assisted design and 2PP micro printing, (ii) sputtering of first a Silica and secondly an Iridium layer, and (iii) activation by dispersion in hydrazine. (B) SEM images of a 3D printed disc, triangle, square, pentagon and hexagon. The dotted red line indicates where the particles were attached to the substrate during printing and sputter coating. All scale bars represent 10 \si{\micro \meter}. (C) Trajectories centered at the origin, recorded over the course of 90s for active polygons in a low density regime and in the presence of N\textsubscript{2}H\textsubscript{4} 0.0005\%v/v. (D) Speed ($U$) and translational diffusion coefficient ($D_t)$ distributions for active polygons. $U$ and $D_t$ were found by fitting the short lag-time dynamics to $MSD= 4D_{0}\delta t+ U^{2}\delta t^{2}$  between $\delta t_{min}= 1s $ and $\delta t_{max}= 5s $. (E) Averaged velocity autocorrelation functions (VACF) at different lag times $\Delta t$ and extracted rotational diffusion constants by fitting the exponential decay of VACF as $\langle \mathbf{e}_{\varphi}(t)\cdot\mathbf{e}_{\varphi}(t+\Delta t)\rangle = e^{-D_{r} \Delta t }$ for all shapes.}
  \label{fgr:fig1}
\end{figure*}

We fabricated active colloidal polygons by (i) 3D micro-printing, (ii) deposition of Ir catalyst and (iii) activation by dispersion of the coated particles in hydrazine~\cite{Gao_2014} as depicted in Fig.~\ref{fgr:fig1}A. The SEM micrographs in Fig.~\ref{fgr:fig1}B show a representative image of the final printed particles for each shape, demonstrating that the intended geometrical symmetries are preserved even at these micrometer length scales.
The polygons self-propel when dispersed in in 0.0005\%v/v hydrazine by catalysis of hydrazine at the Iridium coated side following the gas-forming reaction:
\[
3N_2H_4 \rightarrow N_2 + 4NH_3
\]
\[
3N_2H_4 \rightarrow 2N_2 + 3H_2 + 2NH_3
\]

We extract the speeds and translational diffusion coefficients of the active polygons by fitting the short-time mean-squared displacements. Despite the low hydrazine concentration, the particles were highly active. The mean speeds ranged between $3.15$ \si{\micro \meter\per\second} to $3.83$ \si{\micro \meter\per\second} corresponding to approximately half a body length per second, while individual particles reach speeds as high as $9.00$ \si{\micro \meter\per\second} (Fig.~\ref{fgr:fig1}C, D). The average velocity per shape can be found in $\dagger$ESI. The different particle shapes exhibit similar average speed and qualitatively comparable trajectories. These speeds are also comparable to previously reported speeds for spherical particles propelled by Ir-hydrazine decomposition~\cite{Gao_2014}. Consistent with Gao et al.~\cite{Gao_2014}, we find that activity depends strongly on the hydrazine concentration, with high speeds occurring only in a very narrow concentration range (see Fig.~\ref{fgr:fig3}C.

The average translational diffusion coefficient $D_t$ is similar across all particle shapes, with the mean values lying within two standard deviations of one another. This is consistent with their comparable sizes and three-dimensional geometry, and the identical properties of the surrounding fluid. Although the average values of $D_t$ increase gradually with the number of facets $N$, the differences are small relative to the variability within each population. For comparison,  the Stokes-Einstein relation predicts $D_t=0.049$ \si{\micro\meter\squared\per\second} for a spherical particle in water, which is in good agreement with the measured values for the polygons.

We extract the rotational diffusion $D_r$ for each shape from fitting the decay of the velocity autocorrelation functions to the expected autocorrelation of ABPs. The fitted values correspond to rotational timescales $D_r^{-1}$ of $63$ \si{\second}–$100$ \si{\second}, indicating persistent active motion for all polygons. From the measured single-particle dynamics, we can infer a Peclet number $Pe = \frac{U}{\sigma D_r} \approx 50$, with $\sigma$ the particle radius, 
supporting that the system is in the active regime. The corresponding active contribution to the long-time diffusion, $D_a = \frac{U^2}{2 D_r} \approx 490$ \si{\micro\meter\squared\per\second}, dominates over the thermal one. Activity dominates over diffusion in the MSD for $t > D_t / v_0^2 = 0.004s$, which is well below the temporal resolution of our measurement. Together, these observations justify neglecting translational thermal noise in the SFI and the simulations, Eq.~\ref{eq: SFI displacement EOM} and Eq.~\ref{eq: phoretic displacement EOM}.

Overall, the different polygonal shapes exhibit robust self-propulsion with comparable translational and rotational dynamics, indicating that their rotational symmetry has only a minor effect on their single particle dynamics. This contrasts with reports of shape-dependent propulsion in Zn/SU-8 based microswimmers~\cite{Maric2024}, possibly because of differences in propulsion mechanism and fabrication strategy. 

\subsection{Pair interactions in colloidal polygons}
To investigate pair interactions and their dependence on particle shape, we increased the surface area fraction to $\phi\approx 1\%$ (See $\dagger$ESI Fig. S3). At this packing fraction statistically relevant information on the orientations and distances within the field of view can be obtained, while at the same time all interactions can be considered as pair-wise and instantaneous.

\begin{figure*}[H]
\centering
  \includegraphics[scale=0.7]{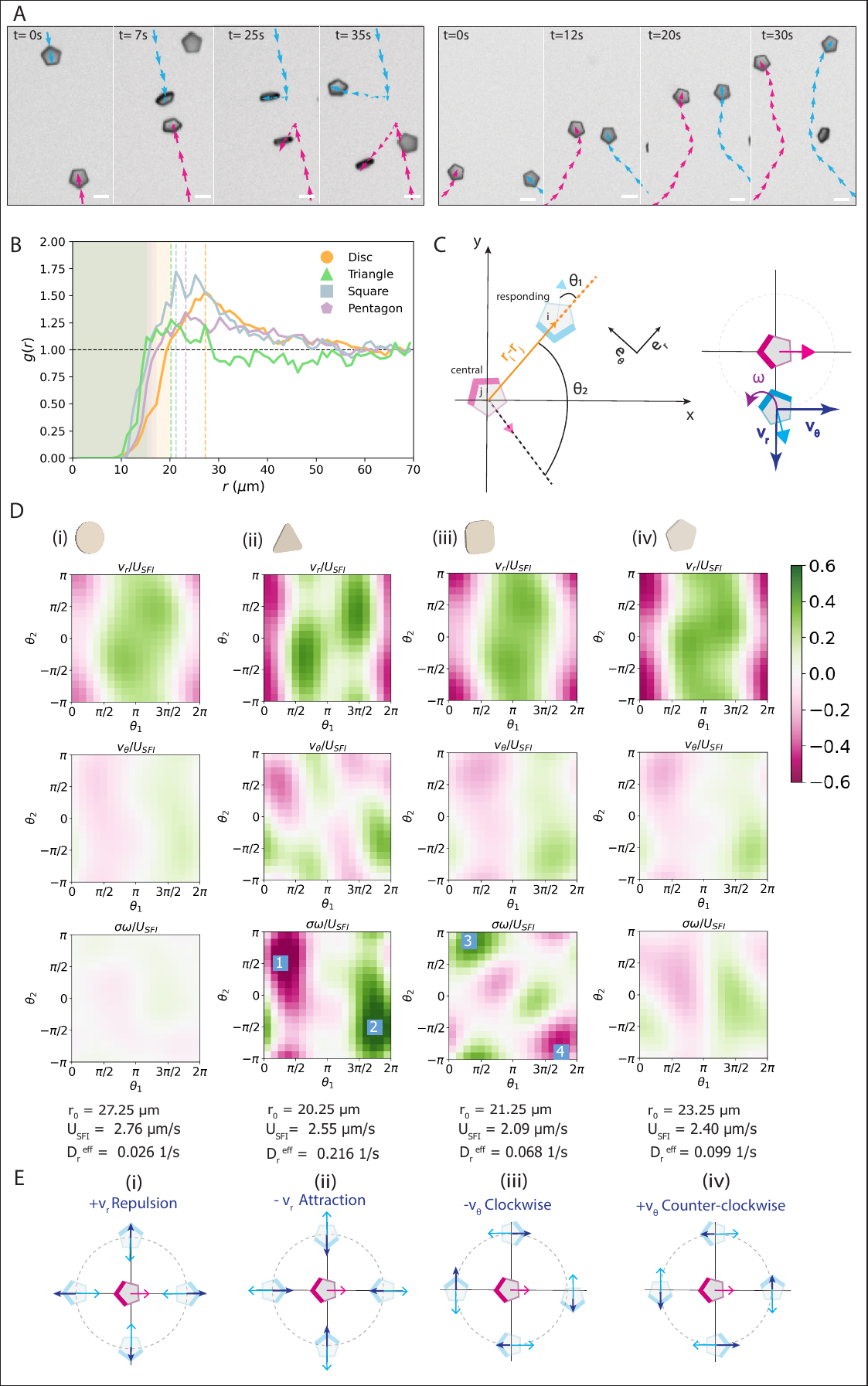}
  \caption{(A) Two brightfield microscopy time series of two interacting particle colliding (left) head-on and (right) side-by-side. The blue and magenta arrows denote the orientation and magnitude of the velocity vectors in time. Scale bars 10 \si{\micro \meter}.(B) Radial pair correlation function $g_l(r)$ for all shapes, vertical line is $r_{0}$ that corresponds to the highest peak per shape. Shaded area corresponds to $r$ values with low statistics. (C) Schematic showing the definition of the angles $\theta_1$ and $\theta_2$ which relate the position and orientation of the central and responding particles. Accompanied by a schematic of the meaning of the radial velocity vector $\mathbf{v_{r}}$, tangential velocity vector $\mathbf{v_{\theta}}$ and angular velocity $\omega$ which are directly obtained from the inference  (D) Magnitude of inferred $\mathbf{v_{r}}$, $\mathbf{v_{\theta}}$ and $\omega$ normalized by the inferred active velocity $U_{\text{SFI}}$, at the relevant distances $r_{0}$ per shape, as a function of $\theta_1$ and $\theta_2$ for (i) discs (ii) triangles (iii) squares (iv) pentagons. Bin size for the angular component is $20^{\circ}$ and for the radial component $\pm 4$ \si{\micro \meter}. (E) Schematic showing the interpretation of the inferred quantities. The interaction effect on the (i-ii) radial and the (iii-iv) tangential component of the velocity is dominated by the orientation $\theta_1$ of the responding particle.The light blue arrows represent the particle's polarity and the dark blue the interaction induced force.}
  \label{fgr:fig2}
\end{figure*}

Qualitative observations of the recorded trajectories revealed that when  particles encounter each other head-on, they decelerate as they approach, reorient, and then resume their characteristic propulsion speed, as illustrated in the left panel of Fig.~\ref{fgr:fig2}A. When the particles interact side by side, they change their direction of motion slightly to avoid each other, as exemplified in the right panel of Fig.~\ref{fgr:fig2}A. The examples shown are for pentagons, but qualitatively similar behavior was observed for all shapes (see Supplementary movie S4).

We quantify these interactions using Stochastic Force Inference (SFI), which reconstructs effective pair-wise interaction velocities and torques directly from particle trajectory data (see Methods). We infer the distance- and orientation-dependent force in the form of an additional effective velocity $\mathbf{v}(r, \theta_1,\theta_2)$ that particle $i$ experiences in the vicinity of particle $j$ (see Equation (1)). 

Here, ${r}$ is the center-to-center distance between the two particles, $\theta_2$ marks the position of the responding particle $i$ relative to central particle $j$ and $\theta_1$ is the orientation of the polarity of the responding particle relative to the bond vector $\mathbf{r}$, as illustrated in Fig.~\ref{fgr:fig2}C. 

The effective velocity $\mathbf{v}$ can be decomposed into a radial contribution $v_r(r, \theta_1,\theta_2)$  and a tangential contribution $v_{\theta}(r, \theta_1,\theta_2)$, with respect to the other particle (see Fig.~\ref{fgr:fig2}C). Please refer to $\dagger$ESI Fig.~S4 for a complete schematic of the interpretation of the relative orientations $\theta_1$ and $\theta_2$. 
SFI can furthermore be used to infer the additional effective angular velocity or torque $\omega (r, \theta_1,\theta_2$), induced by the presence of particle $j$ (Eq. (2)). 

We used the radial pair correlation function $g_1(r)$ to identify the range of separations at which interactions were sampled sufficiently (Fig.~\ref{fgr:fig2}B). The shaded regions correspond to the range of $r$ with low statistics. For each shape, we selected $r_0$ such that it corresponded to the highest peak in $g_1(r)$, and evaluated the angular dependence of the interactions within a radial interval centered on this distance. This choice of $r_0$ captures the strongest interactions at the most statistically relevant distance while preserving orientation-dependent information (see $\dagger$ESI Fig.~S5 and Fig.~S6). Note that the distance at which 
$g_l(r)$ crosses unity moves to larger values with increasing number of edges, because the separation is defined using a center-to-center distance and non-circular particle shape allow closer encounters for particles with fewer edges. 

We applied the SFI algorithm to particles with $N$= 3,4,5 and $\infty$ edges, i.e, discs. For each shape, we obtained the inferred velocity $U_{\text{SFI}}\mathbf{e}_{\varphi_i}(t)$, the  angular diffusion coefficient $D_{r}^{\text{eff}}$, as well as the normalized magnitude of the interaction-induced radial, tangential, and angular velocities $v_r/U_{\text{SFI}}$, $v_\theta/U_{\text{SFI}}$ and $\sigma \omega /U_{\text{SFI}}$. Fig.~\ref{fgr:fig2}D shows the interactions as a function of $\theta_1$ and $\theta_2$ at each of the relevant distances $r_0$ per shape. 

The inferred self-propulsion  speed $U_{\text{SFI}}$ for all particles is in good agreement with the speed measured independently in the dilute regime (see $\dagger$ESI) as shown in Fig.~\ref{fgr:fig2}D (bottom). This is, in turn, a good indication that the SFI algorithm successfully disentangles the self-propulsion  $U_{\text{SFI}}\mathbf{e}_{\varphi_i}(t)$ from the additional interaction velocity $\mathbf{v}$ - Eq.~\ref{eq: SFI displacement EOM}. The inferred rotational diffusion coefficients $D_{r}^{\text{eff}}$ are, on the other hand, larger than those found for the dilute regime, while remaining of the same order of magnitude (Fig.~\ref{fgr:fig2}D, bottom). We attribute this discrepancy to both detection noise (see $\dagger$ESI Fig.~S2) and the presence of higher order effects not captured by the second order SFI. As these contributions are not explicitly included in the inference model, they are, by construction, placed into the stochastic noise term (Equation (\ref{eq: SFI angular EOM})), and therefore we refer to the inferred rotational diffusion coefficient as $D_{r}^{\text{eff}}$. A similar increase has been reported in previous applications of SFI to active particles~\cite{Hem2025}.

Qualitatively, all investigated shapes exhibit similar patterns of attractive and repulsive interactions in both $v_r$ and $v_\theta$ at intermediate interparticle separations ($r\approx2\sigma$), indicating that the dominant interaction symmetry is largely independent of particle shape (Fig.~\ref{fgr:fig2}D). However, differences in the magnitude and angular modulation of these interactions are apparent between shapes, suggesting that the shape influences the strength of the effective interactions without substantially changing their overall structure.

The radial velocity $v_r$ shows both attractive (magenta) and repulsive (green) regions largely independent of the position of the responding particle $\theta_2$, see Fig.~\ref{fgr:fig2}D. Instead the attraction/repulsion is set by its orientation. When the responding particle points away from the central particle (i.e. $\theta_1<\frac{\pi}{2}$ and $\theta_1>\frac{3 \pi}{2}$), it experiences an effective radial attraction, manifested as an decrease in its "outward" radial velocity. This decrease in velocity is largely independent of the value of $\theta_2$, and only slightly stronger when the responding particle is located behind the central particle $-\pi\leq\theta_2<-\frac{\pi}{2}$ or $\frac{\pi}{2}<\theta_2\leq\pi$ (Fig.~\ref{fgr:fig2}D,E(ii)). 
Conversely, when the responding particle points toward the central particle (i.e. $\frac{\pi}{2}<\theta_1<\frac{3\pi}{2}$), the interaction increases the outward radial velocity of the responding particle, indicating an effective radial repulsion. This repulsion is likewise largely independent of $\theta_2$, and only slightly stronger when the responding particle is located on top or below of the central particle, i.e $\theta_2\approx\pm\frac{\pi}{2}$ (Fig.~\ref{fgr:fig2}D,E(i)). Although the repulsive regime occupies approximately the same angular region for all shapes, its magnitude exhibits a more pronounced dependence on $\theta_2$ for triangles and pentagons, particularly near $\theta_1=\pi/2$ and $\theta_1=3\pi/2$. This additional modulation is less apparent for discs, suggesting that geometrical anisotropy contributes to variations in the strength of the repulsion, while leaving the overall interaction pattern largely unchanged. 
Note that electrostatic repulsion cannot account for the strong repulsion observed at the distance probed, as the Debye length only extends over a few nanometers~\cite{Riedel2025}. 

The strong dependence of the radial interaction on $\theta_1$ together with its weak dependence on $\theta_2$ implies that the interactions are non-reciprocal, as exchanging the roles of the central and responding particles does not leave the interaction unchanged. 
We can explicitly write the dominant contribution to the radial force on particle 1 exerted by particle 2 as $\mathbf{v}_r^{12}/U_{\text{SFI}} = - B\cos(\theta_1) \hat{r}_{21}$, where $B>0$ is a dimensionless constant and $\hat{\mathbf r}_{21}$ is the unit vector from particle 2 to particle 1.  As shown in $\dagger$ESI Section 2 and Fig.~S1, in the full coordinate system, this corresponds to $\mathbf{v}_r^{12}/U_{\text{SFI}} = - B\cos(\phi_1-\theta_{21}) \hat{r}_{21}$, with $\theta_{21}$ the polar angle of the bond vector, while swapping labels and orientations leads to $\mathbf{v}_r^{21}/U_{\text{SFI}} =B \cos(\phi_2-\theta_{21}) \hat{r}_{12} \neq - \mathbf{v}_r^{12}/U_{\text{SFI}} $. This experimentally observed effective non-reciprocity is consistent with models of active colloids that consider hydrodynamic and/or phoretic interactions~\cite{Liebchen2021, Fruchart_2026}. 

The weak dependence of $v_r$ on $\theta_2$, together with the remarkably similar interaction patterns observed across all particle shapes, suggests that the interactions are unlikely to originate from conventional pusher–puller hydrodynamics. Such pusher-puller hydrodynamics would be expected to depend strongly on the position of the responding particle with respect to the central one~\cite{Das2016}. Here, instead, the predominantly radial structure of $v_r$ suggests that this quantity is coupled to an approximately radially symmetric concentration gradient generated by the central particle, $\nabla c(\mathbf{r}_i,t)$. The origin and nature of these interactions are investigated further using numerical simulations and discussed in a later section.

The normalized inferred tangential component of the velocity $v_{\theta}/U_{\text{SFI}}$ is smaller compared to the radial component for all polyhedral particles. Yet it remains significant as it can reach $20\%$ of $U_{\text{SFI}}$, see Fig.~\ref{fgr:fig2} D. A purely radial phoretic pull contributes directly to $v_r$, whereas concentration-dependent slowing down contributes to both $v_r$ and $v_{\theta}$. Slowing down alone would give cosine and sine contributions of equal amplitude. The spatial distribution of the chemical field and the additional phoretic pull are therefore essential for understanding the different radial and tangential amplitudes, as discussed in section 4.4 (Interpretation). The inferred tangential velocity is furthermore again only weakly dependent on the positional angle $\theta_2$, and has the same nonreciprocal pattern as the radial component, shifted by $\pi/2$, i.e. $\mathbf{v}_{\theta}^{12}/U_{\text{SFI}} = - B_{\theta}\sin(\theta_1) \hat{e}_{\theta}$, corresponding to clockwise and counterclockwise contributions in the magenta and green regions respectively. As for the radial component, the overall symmetry of the tangential interactions is preserved across all shapes, while their magnitude exhibits a stronger dependence on $\theta_2$ for anisotropic particles than for discs. This suggests that an anisotropic shape introduces additional modulation of the tangential velocity response, without changing its dominant angular structure. 

Finally, the inferred normalized angular velocity, $\sigma \omega/U_{\text{SFI}}$, exhibits a higher magnitude and a rich angular dependence on $\theta_1$ and $\theta_2$ for particles with a triangular and a squared shape. This is a strong indication that geometrical anisotropy is a contributing factor for the appearance of torques for particles interacting at a distance of $r\approx2\sigma$.  Positive and negative values correspond to counterclockwise and clockwise rotation, respectively, and can be interpreted as interaction-induced effective torques on the responding particle (Fig.~\ref{fgr:fig2}D). These torques can be interpreted as aligning or anti-aligning the particle's polarity with the radially outward chemical gradient generated by the central particle. We discuss this further in  section 4.4 (Interpretation). Note that it is possible in the case of the angular velocity $\omega$, to give an estimation of the absolute error. Given the detection error $\varphi_{err}=4^\circ$, the error in $\omega$ becomes $\omega_{err}=4\sqrt{2}/dt\approx18^\circ$ \si{\per\second}. Therefore the strong amplitude peaks in $\omega$ observed for the triangle and the square shapes ($15-25^\circ$ \si{\per\second}) can not be attributed to a noise effect.

\subsection{Phoretic ABP simulations}

\begin{figure*}[H]
\centering
  \includegraphics[scale=0.7]{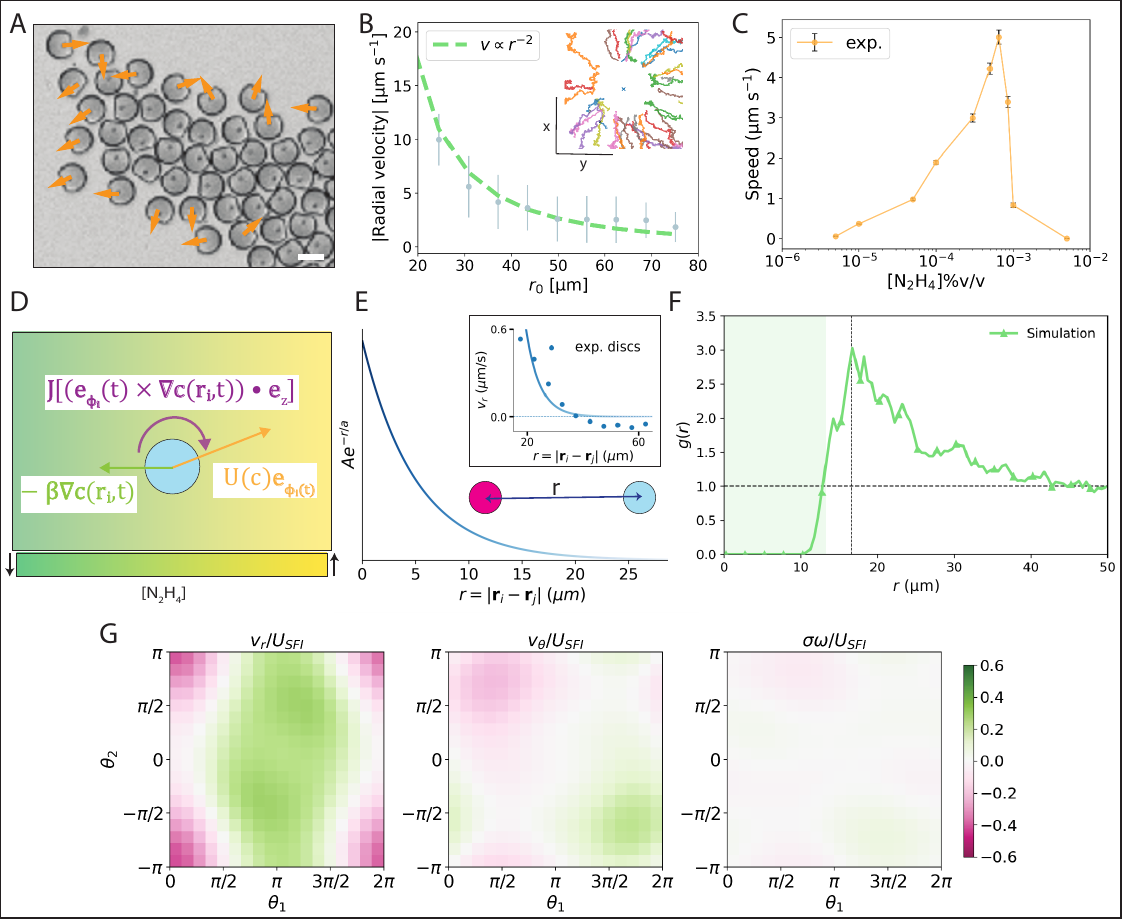}
  \caption{(A) Brightfield snap-shot of experiment performed at higher density ($\phi\approx0.07$) where the orange arrows indicate the polarity of the discs on the periphery. Note that most particles are pointing away from the cluster. (B) Experimentally determined radial velocity of passive tracers around an immobilized active pentagon and $r^{-2}$ fit. Note that the absolute value is displayed and the tracers are attracted to the central particle. Inset: 90 \si{\second} trajectories, the cross represents the center of mass of the fixed polygon. (C) Experimentally determined colloidal speed as a function of hydrazine \%v/v concentration. To capture the strong dependence of the active speed on the concentration in a simple and minimal way, we linearly interpolate the experimental data points on a $log_{10}$-scale. (D) Schematic illustrating the terms included in the numerical simulations for the position and for the orientation (Equations \ref{eq: phoretic displacement EOM} and \ref{eq: phoretic angular EOM}) for particle motion in a concentration gradient. (E) Numerically implemented exponential repulsion between particles $i$ and $j$ such that  $-\sum_{j\neq i}Ae^{-|\mathbf{r}_i - \mathbf{r}_j|/a} \mathbf{e}_{ij}$. Inset: Experimental points recovered from the inferred trigonometric term $\cos([n-|k|]\theta_1+k\theta_2)$ indexed by $n=0$ and $k=0$ along $v_r$ for discs. (F) Radial pair correlation function $g_l(r)$ for simulated data, vertical line is $r_{0}=17$ \si{\micro \meter}. Shaded area corresponds to values of $r$ for which there are low statistics. (G) Inferred normalized pair interactions for data set produced from numerical simulations at inter-particle distance $r_{0}=17$ \si{\micro \meter} as a function of $\theta_1$ and $\theta_2$. Inferred $U_{\text{SFI}}$= 2.56 \si{\micro\meter\per\second} and $D_r^{\text{eff}}$=0.050 \si{\per\second}.}
  \label{fgr:fig3}
\end{figure*}

SFI provides an effective pair-wise description of the particle dynamics but does not identify the physical fields that generate the inferred interactions. To inform the construction of the model further, we experimentally test for the presence of a phoretic attraction, by measuring the radial velocity of tracer particles as a function of distance $r_0$ from an immobilized active particle. We find the typical $\propto r^{-2}$ scaling expected for phoretic attraction~\cite{Liebchen2021}, and a radially isotropic effect where all tracers are pulled equally around the immobilized polygon, see Fig.~\ref{fgr:fig3}B. Motivated by the predominantly radial interaction patterns and the passive-tracer experiment, we therefore test whether a phoretic active-Brownian-particle model can reproduce the experimental features of the interactions. 

The model considers phoresis of a colloidal disc along the chemical gradient of the reaction products, and alignment towards the fuel gradient, as depicted in Fig.~\ref{fgr:fig3}D. This alignment is not only typical in phoretic models, but has also been evidenced in complementary experiments in which, at a higher density, particles form a cluster with a significant fraction of the peripheral particles orienting their polarities along the fuel gradient- as seen in Fig.~\ref{fgr:fig3}A. In addition, the particles self-propel with a fuel-concentration dependent speed (with a maximum speed $U(c^*)$ - Fig.~\ref{fgr:fig3}C), undergo rotational diffusion and experience a repulsion at very short interparticle distances (Fig.~\ref{fgr:fig3}E). We neglect particle geometry in the model to distinguish the common features of the interactions from additional shape-dependent contributions that are not explicitly captured by the model.

All terms were found to be necessary to qualitatively reproduce the experimental inference. In particular, only including the slow-down is insufficient: the off-centered catalytic cap and phoretic pull are also needed to recover the relative radial and tangential magnitudes and the detailed interaction maps. Refer to the $\dagger$ESI Section 8 for implementations of the numerical model in which either the exponential repulsion, the phoretic pull or the off-centered catalytic cap is omitted; in each case, the resulting behavior fails to qualitatively reproduce the experimental observations. 

The precise implementation of the model is described in the Methods section 3.3, see Equations \ref{eq: phoretic displacement EOM} and \ref{eq: phoretic angular EOM}. Simulations were performed using the same duration, sampling rate, and packing fraction as in the experiments (for parameters see $\dagger$ESI Table 1).
The simulated trajectories (see supplementary movies S2 and S3) and orientations were subsequently analyzed using the SFI framework to recover the effective interactions. 

Where possible, model parameters were constrained directly by experimental measurements. We used the measured values of particle size, $D_r$, packing fraction and a direct interpolation from $U(c)$ in Fig.~\ref{fgr:fig3}C, and estimate the effective repulsion strength $A$, phoretic mobility $\beta$ and alignment strength $J$ using successive refinement through the SFI results, together with a best-fit concentration field dynamics. We emphasize that we can not  directly infer the phoretic model parameters from the experiments using SFI, because it would require a reformulation of the SFI algorithm to allow for field mediated interactions, as well as concentration field measurements. Instead, we explore different regimes of the numerical model to investigate whether a phoretic mechanism can semi-quantitatively reproduce the experimentally observed effective pair interactions. This enables us to relate chemically mediated interactions to the resulting effective interactions and compare them directly with the experiments.

We first compare the radial pair correlation function $g_1(r)$ for the data produced by numerical simulations (Fig.~\ref{fgr:fig3}F) with the experimental one. Simulations containing only phoretic coupling and contact interactions did not reproduce the experimentally observed suppression of $g_1(r)$ at close separations ($\dagger$ ESI Fig.~S8). 
We therefore introduced an exponential repulsion as an effective description of additional near-field interactions. This repulsion is most likely of hydrodynamic origin, and becomes relevant as the responding particle approaches the central particle. It is represented in the phoretic model by the term $-\sum_{j\neq i}Ae^{-|\mathbf{r}_i - \mathbf{r}_j|/a} \mathbf{e}_{ij}$ (Fig.~\ref{fgr:fig3}E). With this contribution included, we find good agreement between the experimental and numerical radial distribution function, especially for triangles and squares, for which we observe accumulation at around 20 \si{\micro\meter} followed by a steady decrease until reaching an ideal gas-like distribution at larger distances. Additionally we observe that this form of the repulsion potential corresponds with the inferred zero order term along $v_r$ for discs - see inset of Fig.~\ref{fgr:fig3}E and Eq.~7 in work by one of us (Hem et al.~\cite{Hem2025}).

Applying SFI to the simulated trajectories yields the inferred interactions shown in Fig.~\ref{fgr:fig3}G. By adjusting the balance between fuel depletion, diffusion and relaxation, we obtain a good agreement between the active speed of the particles in simulations, $U_{\text{SFI}}= 2.76$ \si{\micro \meter\per\second}, and experiments (Fig.~\ref{fgr:fig1}D). Furthermore, the extracted rotational diffusion constant $D_r$ is exactly equal to the $D_r$ parameter of the simulation model $0.050$ \si{\per\second}.

The simulations reproduce the dominant angular structure of both the radial and tangential interactions, including their strong dependence on $\theta_1$ and weaker dependence on $\theta_2$ - Fig.~\ref{fgr:fig3}G. Resolving the fuel depletion at the back of the simulated particles additionally captures part of the $\theta_2$ dependence observed experimentally, and provides an excellent match to the SFI results for the discs. However, the more pronounced modulation of the interaction magnitudes for anisotropic particles is not fully reproduced by the geometrically isotropic model. This suggests that shape-dependent differences in the spatial extent and geometry of the catalytic regions, beyond our simplified sampling kernel, contribute to the finer details of the experimentally inferred interactions.

The matched alignment strength $J$ is low, which shows up as a low normalized magnitude of the inferred $\sigma\omega/U_{SFI}$ in Fig.~\ref{fgr:fig3}G, as it also does experimentally for the discs and the pentagons. This further strengthens the argument that the torque observed experimentally for the triangles and the squares originates from their shape anisotropy. This is further discussed in section 4.4 Interpretation.

Finally, we note that we conducted a simulation with a two-fold finer grid size to verify that the chosen grid size was not biasing the inference results ($\dagger$ESI Fig.~S7).

\subsection{Interpretation}

 \begin{figure*}[H]
\centering
  \includegraphics[width=0.8\textwidth]{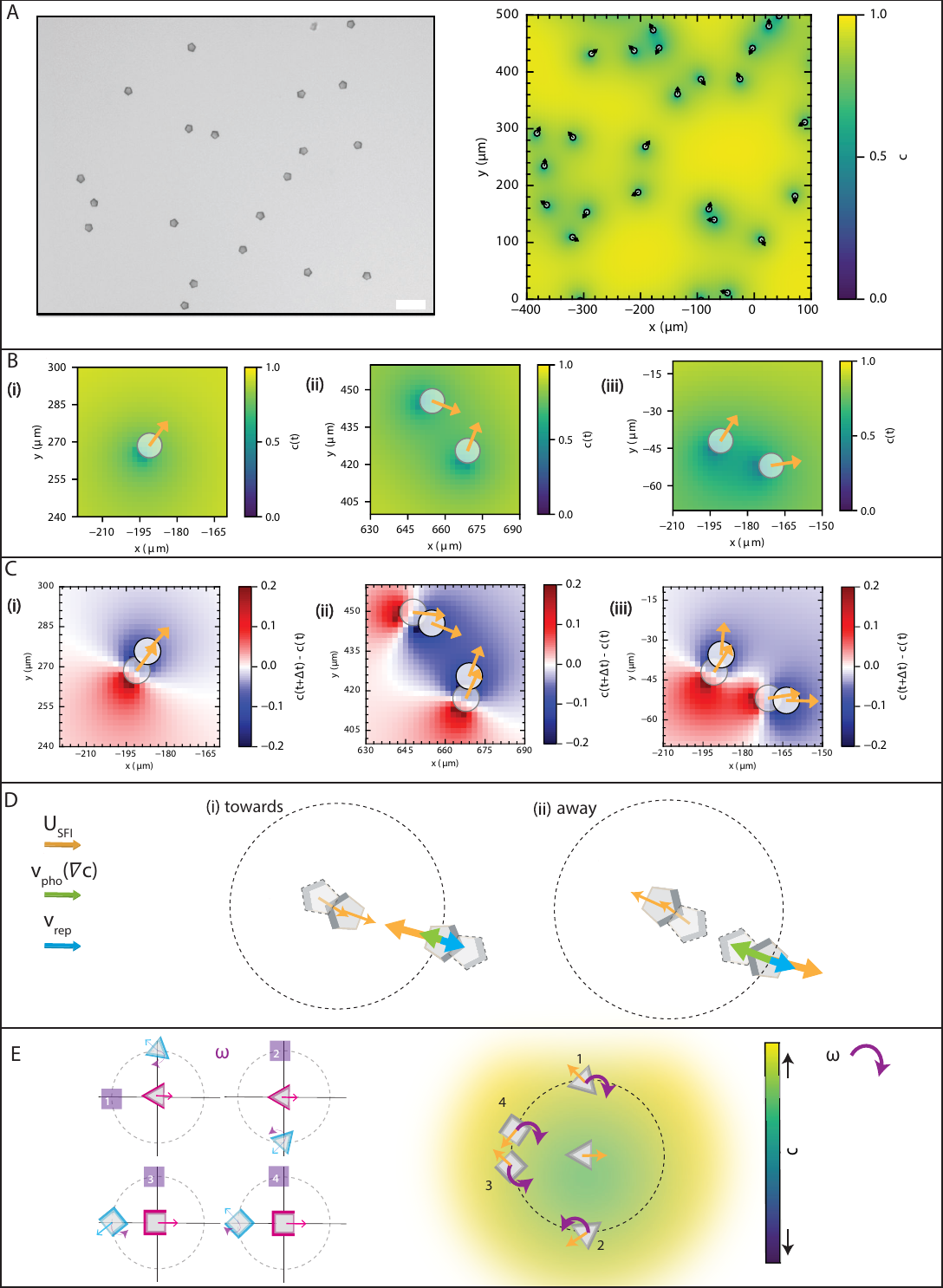}
  \caption{(A) Brightfield microscopy image of pentagons at $\phi=1\%$ (left) accompanied by one frame of numerically simulated point particles at the same area fraction (right). The color bar indicates the magnitude of the chemical field generated around the particles. The field is radial and particles leave no visible trails.  Arrows represent the polarity vectors. The scale bar represents 20\si{\micro\meter}. (B) Instant fuel field magnitude around a single particle (i) and interacting particles (ii)(iii), extracted from two random consecutive frames of the numerical simulation data. The orange arrow depicts the self-propulsion direction. (C) Fuel concentration gradient over times $t+\Delta t$ and $t$ generated by a single particle (i) and interacting particles (ii)(iii), extracted from two random consecutive frames of the numerical simulation data, $\Delta t = 3$s. Blue represents higher fuel concentration at time $t$ with respect to time $t+\Delta t$ and red lower fuel concentration at time $t$ with respect to time $t+\Delta t$. The orange arrow depicts the self-propulsion direction. (D) Illustration of the  different contributions to the particles velocity - the inferred constant velocity, the phoretic pull and the repulsion -  in the most extreme (i) "towards" and (ii) "away" scenarios. The arrows indicate the forces experienced by the responding polygonal particle when interacting with a central particle as deduced from SFI and phoretic tracer experiments. (E) Interpretation of the $\theta_1$ and $\theta_2$ dependent angular velocity $\omega$ experienced by the responding particle assuming a radial $\nabla c $ that extends from the central particle. The numbers indicate the same regions as in Fig.~\ref{fgr:fig2}D(iv).}
  \label{fgr:fig4}
\end{figure*}

The good agreement between the interactions obtained from the  simulation model and experiments provides insights into the origin of the interactions between the active polyhedra. 
We interpret the interaction patterns by first identifying the origin of their common angular symmetry, and then considering how spatial and temporal variations of the chemical field determine their magnitudes and finer structure.

First, by construction, the SFI framework assumes a constant active velocity $U_{\text{SFI}}$. Consequently, any difference between the assumed propulsion speed, $U_{\text{SFI}}$, and the actual propulsion speed, $U(c)$, is interpreted by the inference algorithm as an effective interaction. For a reduction in propulsion speed with an orientation-independent magnitude \(A_{\mathrm{s}}(r)>0\), this contribution takes the form
\[
\mathbf{v}_{\mathrm{slow}}
=-A_{\mathrm{s}}(r)\mathbf{e}_{\varphi_1}
=-A_{\mathrm{s}}(r)\left[
\cos\theta_1\,\mathbf{e}_r+
\sin\theta_1\,\mathbf{e}_\theta
\right].
\]
This contribution explains the dominant angular symmetry of both the radial and tangential interaction maps. The sign of its radial component depends solely on the particle's orientation angle $\theta_1$: when the responding particle is oriented away from the central particle ($\theta_1<\frac{\pi}{2}$ or $\theta_1>\frac{3 \pi}{2}$), a slower velocity appears as an effective attraction. Conversely, when the particle is oriented towards the central particle ($\frac{\pi}{2}<\theta_1<\frac{3\pi}{2}$), the same reduction in propulsion speed is inferred as an effective repulsion. 

Second, the phoretic pull and off-centered catalytic side are essential for recovering the relative radial and tangential amplitudes and the detailed interaction maps. The magnitude of the effective attraction and repulsion depends on both the local fuel concentration, through $U(c)$, and its gradient, through the phoretic pull.

As mentioned before, slowing down alone would give radial cosine and tangential sine contributions of equal amplitude. However this is not the case  for the interaction maps obtained for the experimental and simulated data in Fig. \ref{fgr:fig2}D and Fig. \ref{fgr:fig3}G. For both experiments and simulations, in the attractive region, the weak dependence on $\theta_2$ is consistent with a phoretic pull stemming from a mostly radial fuel concentration gradient generated by the responding and central particles, which we confirmed with passive tracer experiments (Fig.~\ref{fgr:fig3}B). Note that the specific relative orientation of the responding and central particles modulates the strength of the interactions as observed by the change in color intensities in Fig.~\ref{fgr:fig2}D and and Fig. \ref{fgr:fig3}G. Also note that since the diffusion of the field is fast compared to the speed of the particles, the depletion zone is not a trail left behind the particle, but instead a radial profile around the particle that moves with it (Fig.~\ref{fgr:fig4}A, B(i) and Supplementary movie S2).

The extended size of the particles and the partial coating with the catalyst possibly further explain the variations in $\theta_2$. Given that the particles have a partial catalytic coating in the back, the local fuel gradient is modulated by the distance between those catalytic sites as we can recover from simulation data in Fig.~\ref{fgr:fig4} B(ii),(iii). At the back of the central particle the catalytic side depletes the fuel, and so is the location of the steepest gradient. Hence the inferred radial attraction is stronger when the central particle is pointing with its cap to the responding particle ($\theta_2=\pm\pi$), and strongest when the responding particle is pointing with its cap towards the central particle as well ($\theta_1=0=2\pi$) Fig.~\ref{fgr:fig4}D(ii). As mentioned earlier, we can only capture this effect in simulations by resolving the field interactions at the back of the particle (Methods section 3.3.).
Note that this effect is more pronounced for anisotropic particles in experiments than for discs or simulations, suggesting that geometrical anisotropy plays an important role in modulating its magnitude. One possible origin is the shape and spatial extent of the catalytic regions, which vary between particle geometries. Even at the same center-to-center separation, these differences can change the relative positions of the catalytic surfaces by several micrometers, thereby modifying the local concentration gradients experienced by the particles. Given the strong dependence of both phoretic interactions and propulsion speed on the local chemical environment, such geometrical differences could account for the enhanced variation in interaction magnitude without changing the dominant attractive and repulsive regimes. 

Third, the particles can further modify these spatial effects through the temporal evolution of the fuel field. This effect is visualized by the simulated fuel field differentials in Fig.~\ref{fgr:fig4}C, where blue denotes an environment that starts to be depleted at time $t+\Delta t$ with respect to time $t$ and red an area that is recovering from the particle's passing. When the responding and central particles move towards each other, they just start to deplete the region in between (Fig.~\ref{fgr:fig4}C(ii)) and therefore the phoretic pull is weak (Fig.~\ref{fgr:fig4}C(ii)). Conversely, when the responding and central particle are moving away from each other, their catalytic ends have previously depleted the fuel together, thereby creating a steeper gradient, as seen in red in Fig.~\ref{fgr:fig4}C(iii) where the region in between starts to recover. The responding particle consequently experiences a larger phoretic attraction as depicted in Fig.~\ref{fgr:fig4}D(iii). The spatial arrangement of the catalytic regions can amplify the effect of these temporal concentration differences on the local propulsion speed and phoretic pull, further modulating the interaction magnitudes.

Additionally, the repulsive regime in $v_r$ contains a short-range repulsive contribution. Initial simulations without this term reproduced the correct interaction symmetry, i.e. stronger attraction for $\theta_1<\frac{\pi}{2}$ and $\theta_1>\frac{3 \pi}{2}$, with little or no attraction elsewhere, but they failed to reproduce the characteristic repulsive band observed experimentally (see †ESI Fig.~S8). Since this contribution is independent of the particle orientations for a fixed particle separation, it appears as a constant offset (compare Fig.~\ref{fgr:fig3}G with Fig.~S8). 

To summarize this interpretation, Fig~\ref{fgr:fig4}D presents a schematic of the different contributions to the particles velocity - the inferred constant velocity, the phoretic pull and the repulsion - in the most extreme scenarios.\\

As the responding particle gets closer to the central particle, the repulsion and attraction become stronger in line with the radial dependency of the phoretic pull (Fig.~S6). However, we refrain from drawing strong conclusions at these shorter distances, as the available statistics are limited, the tracking becomes less reliable and particles occasionally undergo flips, i.e rotations in the $z$ direction. Refer to section 3.2 Stochastic Force Inference for the discussion on the uncertainty at these distances. On the other hand, at larger separations, the data provide a clear picture of the spatial extent of the chemical field. Irrespective of particle shape, the gradient appears to remain detectable up to at least $r_0=40$ \si{\micro \meter} from the center of mass of the central particle ($\dagger$ESI Fig.~S6). In all cases, the self-propulsion speed $U_{\text{SFI}}$ of the responding particle remains sufficiently large to overcome these interactions.\\

These observations indicate that concentration dependent slowing down sets the dominant angular symmetry of the translational interactions, while the spatial distribution and temporal evolution of the chemical field, together with phoretic attraction, determine their relative magnitudes and finer angular dependence. This explains why the same attractive and repulsive regimes are recovered for all experimental shapes and  the geometrically isotropic model. Nevertheless, the differences in interaction magnitude observed within these regions suggest that particle geometry can modify the local chemical environment and, consequently, the strength of the effective interactions, without substantially changing their overall symmetry.\\

The tangential interaction follows from the tangential projection of the same local concentration-dependent slowdown that contributes to the radial response. Its magnitude is modulated by the spatial distribution of the fuel concentration, while the predominantly radial phoretic pull contributes mainly to $v_r$.  
Again, the decomposition of the fuel by the responding and central particle add up, leading to an additional difference in the local gradient caused by the responding particle. The responding particle thus consistently experiences an effective tangential pull in  the direction opposite of its polarity vector, see Fig.~\ref{fgr:fig2}E(iii) and E(iv), largely independent of the particle shape. In the experimental data, we additionally see a stronger response of the responding particle  at the back of the central particle ($\theta_2=\pm\pi$) probably because the catalytic site there leads to stronger local fuel depletion. The larger magnitude observed for anisotropic particles may once again be attributed to variations in the shape and extent of their catalytic sites. The simulations also reproduce this effect, albeit to a lesser extent (Fig.~\ref{fgr:fig3}G), exhibiting a response more closely resembling that observed for discs.\\

Finally, we consider the induced angular velocity $\omega$. The alignment term included in the simulations rotates the particle polarity toward the gradient ($\nabla c$) that increases outwards from the central particle (Fig.~\ref{fgr:fig4}A (right panel) and Fig.~\ref{fgr:fig3}A). The value given to the alignment strength in simulations, $J$, is small and results in very small torques for the point particles. However we see this gradient alignment appear strongly in the effective torques inferred for the experiments for the triangles and, more weakly, for the pentagons. This effect shows up as a clock wise rotation (magenta) when $0<\theta_1<\frac{\pi}{2}$ and counter-clock wise (green) when $\frac{3\pi}{2}<\theta_1<2\pi$ (scenarios 1 and 2 in Fig.~\ref{fgr:fig2}D and Fig.~\ref{fgr:fig4}D).
Increasing $J$ in simulations, also leads to this banded structure (see Fig.~S10), so it may be that the value of $J$ depends on the particle shape. The presence of this type of repulsive torques in extended anisotropic particles and its absence in isotropic ones and point particles, strongly suggests their origin being alignment to the gradient strengthened by geometrical anisotropy.

In experiments, anti-alignment is observed also for active squares. We can observe this effect when the responding particle points approximately away from the central particle, i.e. scenarios 3 and 4 in Fig.~\ref{fgr:fig2}D and Fig.~\ref{fgr:fig4}D. In these regions, the particle is rotated towards the central particle and thus in anti-alignment with the gradient. 
This effective interaction is not reproduced by the numerical model and not present in other shapes, suggesting that it does not originate from the gradient alignment but from other physical mechanisms not captured by our plausible description. 
The particles shape may induce additional hydrodynamic or phoretic effects that become relevant under experimental conditions and can give rise to complex torques. Their microscopic origin, however, remains an open question. 

\section{Conclusions}
We introduced an experimental platform for investigating the single-particle dynamics and pair interactions of active colloidal polygons with systematically varied rotational symmetry. The colloids were fabricated using 3D microprinting and propelled through the catalytic decomposition of hydrazine at an iridium layer deposited on one of their sides. In the dilute regime, all particle shapes exhibited comparable propulsion speeds of approximately half a body length per second towards their uncoated side, as well as similar rotational persistence ($\tau_r\approx 100$ \si{\second}), indicating that particle shape has only a minor influence on their single-particle dynamics.

Using Stochastic Force Inference, we inferred the effective pairwise interactions as a function of the relative positions, orientations, and interparticle distances of the particles. The translational interactions were predominantly radial and exhibited similar attractive and repulsive regimes across all shapes, governed primarily by the orientation of the responding particle. However, within these common regimes, the interaction magnitudes showed additional shape-dependent variations. The interactions were intrinsically non-reciprocal, as exchanging the particle identities did not generally reverse the interaction-induced velocities. Their spatial range and common angular structure suggest that long-range chemical concentration fields dominate over direct steric interactions at the probed distances.

A minimal plausible phoretic ABP model, incorporating concentration field dynamics, concentration-dependent propulsion speed, and finite-range repulsion, successfully reproduced the dominant symmetries of the experimentally inferred interactions. This agreement supports chemically mediated interactions as the main origin of the common patterns of non-reciprocal attraction and repulsion across particle geometries. Concentration-dependent slowing sets the common angular symmetry, while the spatial geometry of the chemical coupling and phoretic attraction are essential for recovering the relative radial and tangential magnitudes and detailed interaction maps, even without explicitly representing particle shape. However, the stronger modulation of the interaction magnitudes observed for anisotropic particles was not captured by the model, suggesting that geometrical differences, potentially through the shape and spatial extent of the catalytic regions, influence the local concentration gradients. Furthermore, the model did not reproduce the pronounced alignment and anti-alignment responses observed experimentally for triangles and squares, pointing to additional contributions from geometrical anisotropy, hydrodynamic interactions, and finite-size effects. Thus, particle shape modulates the strength and orientational features of the interactions without fundamentally altering their dominant structure.

Our work studies one of the most central topics in active colloids science: quantification of pair-wise interactions and their physical origin. We demonstrated that a combination of SFI, experiments and numerical simulations provides a tool to extract and interpret the pair interactions of active colloids. This approach can be extended to other particle geometries and activation mechanisms as well as biological active matter, thereby providing the basic information needed to understand and from there design active matter systems. 

\printcredits

\section*{Conflicts of interest}

There are no conflicts to declare.

\section*{Data availability}

Data is available at: https://doi.org/10.4121/c47856fd-260c-4408-aa8f-5691ad2877f1.

\section*{Acknowledgments}
This work was supported by The Dutch Research Organization  (NWO/OCW), as part of the Vidi scheme (grant no. VI.Vidi.193.069). The authors thank Rachel Doherty for help with 3D printing and SEM imaging, and Alexandre Morin for useful discussions. DJK gratefully acknowledges ESPCI Paris for hosting her as a visiting professor, which initiated discussions with Vincent Démery and the collaboration from which this work came forth.

%\bibliographystyle{cas-model2-names}
\bibliographystyle{unsrtnat}
\bibliography{Active_Polygons}

\end{document}

\maketitle

\clearpage
\textbf{Supplementary movie S1:}
Brightfield microscopy images of active polygon colloids at $\phi \approx 1\%$ including trajectories in time (blue). Display speed is 20$\times$ real time.

\textbf{Supplementary movie S2:}
Video of the phoretic model simulation with grid spacing $l_{\mathrm{bin}}=2$ \si{\micro\meter}. Visible are the value of $c$, the location of the particles and their polarity vectors.

\textbf{Supplementary movie S3:}
Video of the phoretic model simulation with grid spacing $l_{\mathrm{bin}}=2$ \si{\micro\meter}. Visible are the value of $c$, the location of the particles, their polarity vectors, as well as the trajectories of the particles over time.

\textbf{Supplementary movie S4:}
Exemplary movies of active polygon colloids interacting. The vectors represent the velocity. Display speed is 10$\times$ real time.

\clearpage

\section{Simulation parameters}
We work in units where the friction coefficient $\zeta=1$ and angular friction coefficient is $\zeta_r=1$, and, due to rescaling of the field outlined in the previous section, $c_0=1$. The other parameters are of the values and dimensions shown in table \ref{tbl:simulation_parameters}. We considered both a grid spacing $l_{\mathrm{bin}}=2$ \si{\micro\meter} (a fifth the size of a particle) and $l_{\mathrm{bin}}=1$ \si{\micro\meter} (a tenth of the size of a particle) with identical results.
\begin{table}[h]
\centering
\small
  \caption{\ Parameters used in the simulation}
  \label{tbl:simulation_parameters}
  \begin{tabular*}{0.48\textwidth}{@{\extracolsep{\fill}}lll}
    \hline
    Parameter & Value\\
    \hline
    Particle diameter  & 10 \si{\micro\meter}  \\
    Field grid spacing $l_{\mathrm{bin}}$ & 1-2 \si{\micro\meter}  \\
    System size $L^2$ & $1500 \times 1500$ \si{\micro\meter\squared}  \\
    Packing fraction $\phi$ & $1\%$ \\
    Repulsion amplitude $A$ & 30 \si{\micro\meter\per\second} \\
    Repulsion length scale $a$ &  5 \si{\micro\meter}\\
    Rotational diffusion coefficient $D_r$ &  $0.05$ \si{\per\second} \\

    Field depletion rate $k_d$ &  300 \si{\micro\meter\squared\per\second}\\
    Field diffusion constant $D_c$ &  400 \si{\micro\meter\squared\per\second}\\
    Field relaxation rate $k_r$ &  0.3 \si{\per\second}\\
    Alignment strength $J$ &  2 \si{\micro\meter\per\second}\\
    Phoretic mobility $\beta$ &  120 \si{\micro\meter\squared\per\second}\\
    \hline
  \end{tabular*}
\end{table}
\newpage

\section{Non-reciprocity of the effective radial pair force}
\begin{figure}[h]
\centering
  \includegraphics{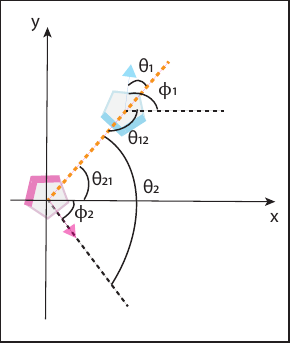}
  \caption{Coordinate system relating $\theta_1$ and $\theta_2$ from SFI to the position and polarities of the responding and central particles in terms of $\phi_1$ and $\phi_2$.}
  \label{fgr:fig2SI}
\end{figure}

Here we show explicitly that the inferred radial force is non-reciprocal. To show this, we first write a functional form for $v_r/U$ that has the same symmetry as the pink-green structure in Figure 2 D(i). Based on the pattern we see there, we consider $\boldsymbol{v}_r^{1 2} \sim -\cos(\theta_1) \boldsymbol{\hat{r}_{21}}$, where we use the superscript $12$ to denote that it is the force on particle 1 (the responding particle) by particle 2 and where $\boldsymbol{\hat{r}_{21}}$ is the unit vector pointing from particle 2 to particle 1. We first express $\boldsymbol{v}_r^{1 2}$ in terms of the polarity angles of the two particles, then swap the particle labels to get the force on particle 2 by particle 1. So, we write $\theta_1 = \phi_1 -  \theta_{21}$ , with $\theta_{21}$ the polar angle of $\boldsymbol{\hat{r}_{21}}$ (Fig.~\ref{fgr:fig2SI}), so that $\boldsymbol{v}_r^{1 2} \sim -\cos(\phi_1 - \theta_{21}) \boldsymbol{\hat{r}_{21}}$. If we now swap particle labels, we find $\boldsymbol{v}_r^{2 1} \sim -\cos(\phi_2 - \theta_{12} ) \boldsymbol{\hat{r}_{12}}$. Using that $\theta_{12} = \theta_{21} - \pi  $, we obtain $\boldsymbol{v}_r^{2 1} \sim -\cos(\phi_2 + \pi - \theta_{21}) \boldsymbol{\hat{r}_{12}} = \cos(\phi_2 - \theta_{21})\boldsymbol{\hat{r}_{12}}$. Since $\phi_1$, $\phi_2$, and $\theta_{12}$ are all independent, $\boldsymbol{v}_r^{1 2}\neq  - \boldsymbol{v}_r^{21}$ and so the radial force is clearly non-reciprocal. We can also express this in the original coordinate system. Noting that $\theta_2 = \theta_{21} - \phi_2$, we arrive at $\boldsymbol{v}_r^{2 1} \sim -\cos(\pi-\theta_{2}) \boldsymbol{\hat{r}_{12}} =\cos(\theta_2) \boldsymbol{\hat{r}_{12}}$. Again, $\boldsymbol{v}_r^{1 2}\neq  - \boldsymbol{v}_r^{21}$.   

\section{Average active velocity $U$ of polygons in the diluted regime}

\begin{table}[h]
\centering
\small
  \caption{\ Average active velocities in diluted regime}
  \label{tbl:velocitites}
  \begin{tabular*}{0.48\textwidth}{@{\extracolsep{\fill}}lll}
    \hline
    Shape & $U$(\si{\micro\meter\per\second})\\
    \hline
    Disks  & 3.776 $\pm$ 1.796\\
    Triangles  & 3.175 $\pm$ 1.533\\
    Squares  & 3.152  $\pm$ 1.111\\
    Pentagons  & 3.832 $\pm$ 2.036\\
    Hexagons  & 3.705 $\pm$ 2.030\\
    \hline
  \end{tabular*}
\end{table}
\newpage
\section{Center of mass and angle determination for SFI from microscopy images}

\begin{figure}[H]
\centering
  \includegraphics{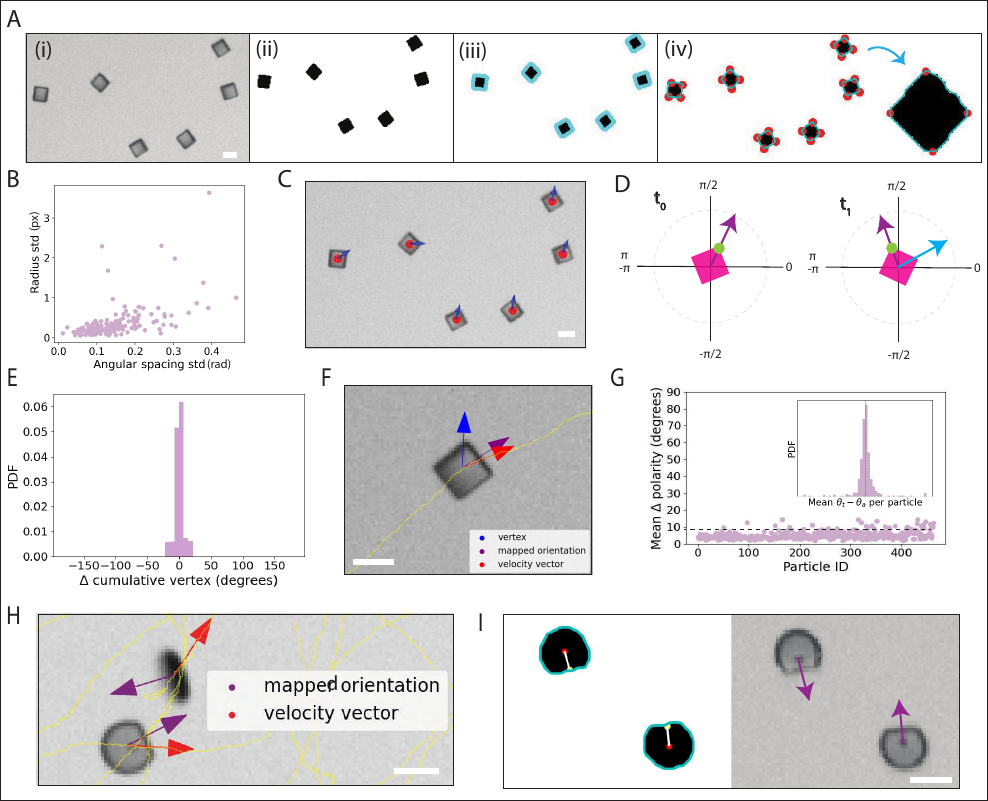}
  \caption{(A) Illustrative panel showing for squares (i) raw microscopy image (ii) binarization (iii) contour detection an (iv) vertex detection. (B) Measure of the deviation of the detected vertices in a random frame from the radial and angular spacing conditions for ideal squares. (C) Orientation detection as the firt vertex from the horizontal. (D) Schematic of the accumulation of the orientation over time. (E) The distribution of the change of
the cumulative vertex over time for squares among all 6000 frames. (F) Example of the mapping of the detected and accumulated orientation into the meaningful polarity referred as mapped orientation. The polarity of the velocity vector is also included for reference. (G) Change in polarity over frames for squares. The dotted horizontal line represents $10^{\circ}$. The inset represents the same measure but per particle. (H) Example of polarity tracking where he shape is no longer a polygon in 2D. (I) Example of polarity tracking in discs.} 
  \label{fgr:fig3SI}
\end{figure}

The center of mass and angle determination from 10x Brightfield microscopy images started with an image enhancement algorithm that normalized the frame intensities, applied a Gaussian filter and binarized the images based on Otsu thresholding. This allowed for accurate identification of the features against the background (see Fig.~\ref{fgr:fig3SI} panel A(i),(ii)). The binary images were then processed to find a smooth contour around each feature (see Fig.~\ref{fgr:fig3SI} panel A(iii)) that allowed for float pixel positions to avoid the "stair-case" effect due to poor resolution. From the contour positions, the curvature ($\kappa$) of each contour is computed as the tangent angle ($\theta$) to the change of position over the contour ($s$) length, and averaged using a Gaussina filter ($G$) to capture real changes and not just pixel spikes:

\begin{equation}
\theta = \arctan2\left(\frac{dy}{ds}, \frac{dx}{ds}\right)
\end{equation}

\begin{equation}
\kappa \approx \left| \frac{d\theta}{ds} \right|
\end{equation}

\begin{equation}
\kappa_i^{(s)} = \sum_j \kappa_j \, G(i-j).
\end{equation}

The points $x_i$ and $y_i$ resulting from the highest $\kappa_i^{(s)}$ are taken to be the vertices of the polygon. To avoid vertex misidentification, the points are turned into polar coordinates using the center of mass ($c$) and compared to ideal regular polygon angles ($\theta_k^{\text{target}}$) as shown in Equation (\ref{eq:eqSI1}):

\begin{equation}
\theta_i = \arctan2(y_i - c_y,\; x_i - c_x)
\end{equation}

\begin{equation}
\theta_k^{\text{target}} = \phi + \frac{2\pi k}{N}
\end{equation}

\begin{equation}
\Delta \theta = \left| \arg\left(e^{i(\theta_i - \theta_k)}\right) \right|,
\label{eq:eqSI1}
\end{equation}

where $\phi$ is an angle to correct for the polygon rotation, $N$ is the symmetry and $k$ is the vertex index. Then, a scoring function is used to choose the most accurate vertices by weighting two different terms: (i) the curvature strength of the candidates (ii) the missmatch to the ideal polygon angle (with $\lambda=5$), as shown in Equation (\ref{eq:eqSI2}):

\begin{equation}
\text{score} = \sum_{i=1}^{N} \left( \kappa_i^{(s)} - \lambda \, \Delta\theta_i \right).
\label{eq:eqSI2}
\end{equation}

Fig.~\ref{fgr:fig3SI} panel A(iv) depicts an example of the detected vertices in red. Panel B shows a measure of the accuracy of the detected vertices for an aleatory frame with 100+ squares by comparing the standard deviation of the angular spacing (in the ideal case all vertices should be spaced regularly) and the standard deviation of the radius at which the vertices fall (ideally all vertices should be on the same radius from the center of mass). \\

Once the vertices of the polygons are detected the next step is to track a meaningful inherent polarity over time. For this we first label individual particles using TrackPy. The original orientation that comes out from our vertex detection is always the first vertex with respect to the horizontal as shown in Fig.~\ref{fgr:fig3SI}C. To keep vertex detection consistent across frames, it is important to accumulate the angle in time based on the symmetry, this means to always find the shorter angular distance between the detected angle at $t_0$ and at $t_1$ as shown in Fig.~\ref{fgr:fig3SI} D. The distribution of the change of the cumulative vertex over time for squares among all 6000 frames is shown in Fig.~\ref{fgr:fig3SI} E. To go from a vertex to a meaningful polarity, the accumulated orientation is then mapped to the active vector for the first 20 frames of each particle simply by adding up an offset angle, to represent the active motion polarity. To not skew the inference results this is only done once, for all other frames the same offset is used to map the detected vertex onto the polarity. An additional condition on the rotational speed and the position of the polarity is implemented such that polarity cant change from frame to frame more than $30^\circ$ and such that it always lays on the middle of a flat edge. This resulted in the mapping that is exemplified in Fig.~\ref{fgr:fig3SI} F. Finally particles that still showed spikes (change in the polarity over frames more than $10^\circ$) were removed from the final data frame as showed by Fig.~\ref{fgr:fig3SI} G. The standard error for the detection of the orientation is estimated to $\varphi_{err}\approx 4^\circ$. This value is obtained by fitting the mean squared angular displacement (MSAD) with $\left\langle \left[\varphi(t+\Delta t)-\varphi(t)\right]^2 \right\rangle = 2D_r\Delta t + 2\varphi_{\mathrm{err}}^2$. 

\paragraph{Special treatment for particles that leave the x,y plane.} Sometimes particles loose their 2D shape and "stand-up" on the substrate as exemplified in the main text Figure 2A(i). This is a phenomenon already discussed in the main which is most likely linked with hydrodynamic effects. Since the polygonal shape is no longer present, a different strategy was followed to keep track of the particles' polarity. For this we established an area threshold per shape where the polygon was no longer distinguishable, and fit an ellipse instead. The polarity was set to be the perpendicular vector to the long axis of the ellipse that correlated the most with the particle's previous orientation, as can be observed in panel H.

\paragraph{Special treatment for discs.} Due to the absence of vertices in discs, the detection of the polarity was treated differently. In this case we use a straight line detection to accurately track the flatter side (cutoff) of the discs, and the polarity is set to be the vector between the center of mass and the middle of the detected straight line, see Fig.~\ref{fgr:fig3SI} I. This means no mapping is necessary and the area correction stays the same as for the other polygons. 

\newpage

\section{Microscopy observations- Fields of view}

\begin{figure}[H]
\centering
  \includegraphics{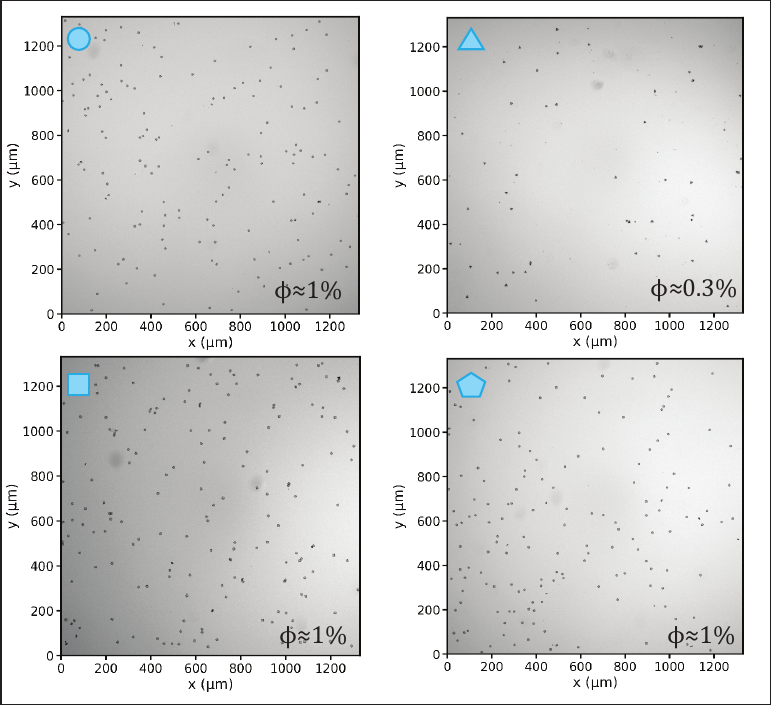}
  \caption{Fields of view as observed by Bright Field microscopy using a 10x air objective and indicating the estimated packing fraction $\phi$ per shape.}
  \label{fgr:fig4SI}
\end{figure}

\newpage

\section{Guide for reading orientations}

\begin{figure}[H]
\centering
  \includegraphics{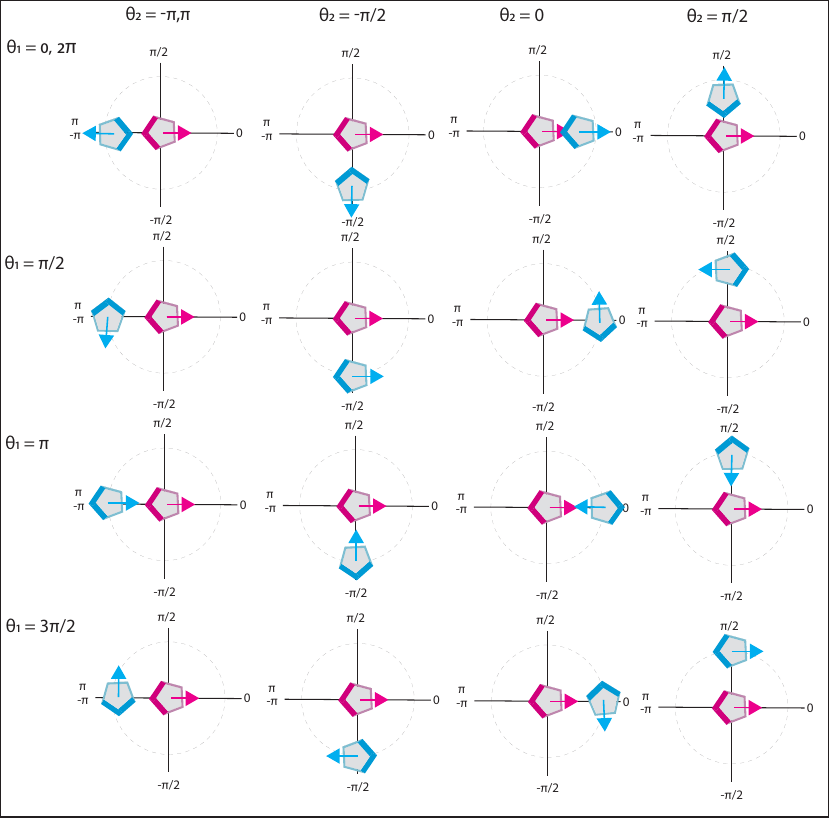}
  \caption{Compilation of visual representation for different combinations of $\theta_1$, the orientation between the active force director of the responding particle and the bond vector, and $\theta_2$, the position of the responding particle on the orbit of the central particle. }
  \label{fgr:fig5SI}
\end{figure}

\newpage

\section{Additional inference results}

\subsection{Histogram indicating the orientation dependent pair correlation function for different interparticle distances - Experiments and Simulations}

\begin{figure}[h]
\centering
  \includegraphics{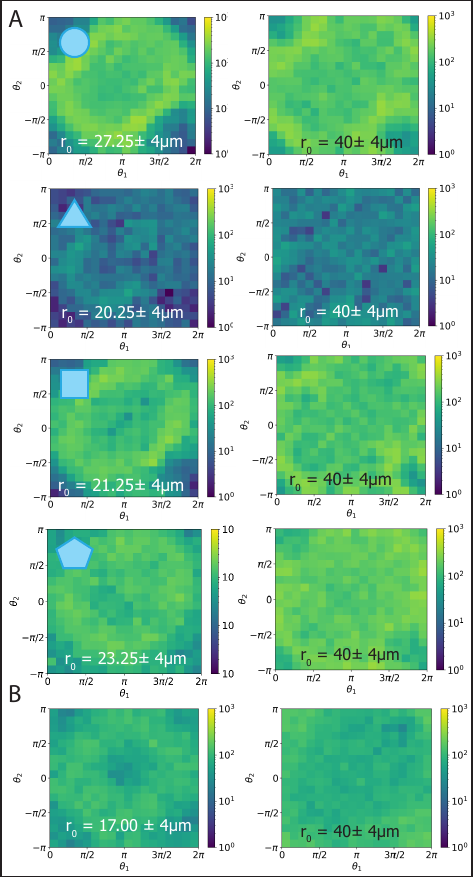}
  \caption{Histograms indicating the orientation dependent pair correlation function for relevant interparticle distances for (A) experiments and (B) simulations. Bin size for the angular component is $20^{\circ}$ and for the radial component $\pm 4\mu m$.}
  \label{fgr:fig6SI}
\end{figure}

\subsection{Distance dependent behavior of inferred quantities - Experiments and Simulation}

\begin{figure}[H]
\centering
  \includegraphics{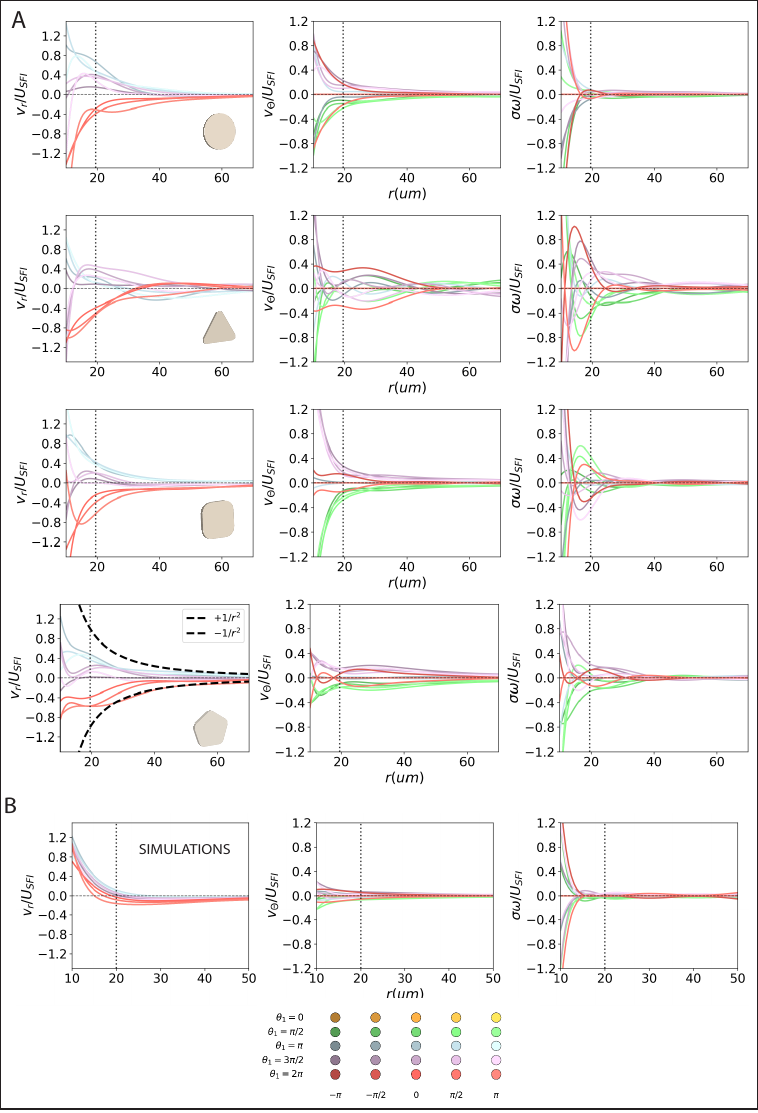}
  \caption{Normalized $v_{r}/U_{SFI}$, $v_{\theta}/U_{SFI}$ and $\omega/U_{SFI}$ as a function of the radial distance
  $r$ for different combinations of the angles $\theta_1$ and $\theta_1$ extracted by SFI from data generated by (A) experiments and (B) numerical simulations. The angular combinations are indicated in color as defined at the bottom of the figure. The vertical line marks $r_0=2\sigma$}
  \label{fgr:fig7SI}
\end{figure}
\newpage

\section{Additional simulations}

\subsection{Finer grid simulation}

\begin{figure}[h]
\centering
  \includegraphics{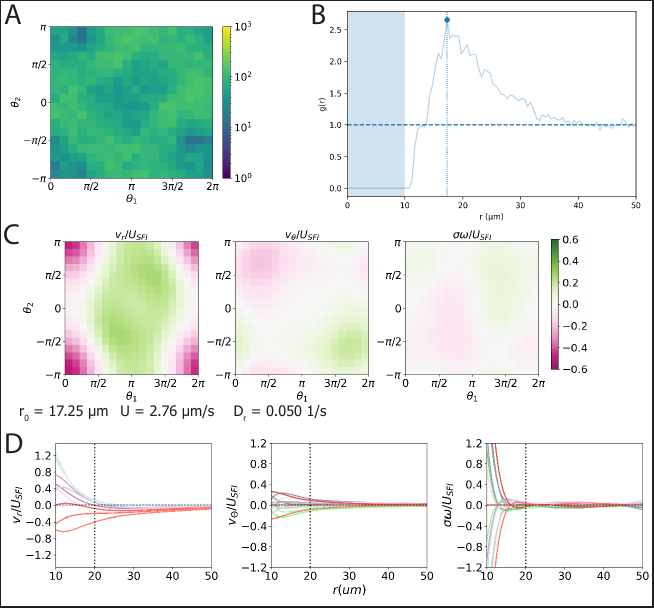}
  \caption{Inference results for a simulation grid with 5 fold smaller grid distance, \textit{i.e.}~$l_{\text{bin}} = 1$ \si{\micro\meter}. (A) Pair correlation function for all relative orientations at $17.25$ \si{\micro\meter} (B) Pair correlation function over the cell length (C) Inferred normalized pair interactions inter-particle distance $r_{0}=17.25$ \si{\micro \meter} as a function of $\theta_1$ and $\theta_2$. (D) Normalized $v_{r}$, $v_{\theta}$ and $\omega$ as a function of the radial distance for different angle combinations. The vertical line marks $r_0=17.25 \mu m$. The color code is the same as used in Figure S7. }
  \label{fgr:fig8SI}
\end{figure}
\clearpage
\subsection{Simulation without exponential repulsion}
\begin{figure}[h]
\centering
  \includegraphics{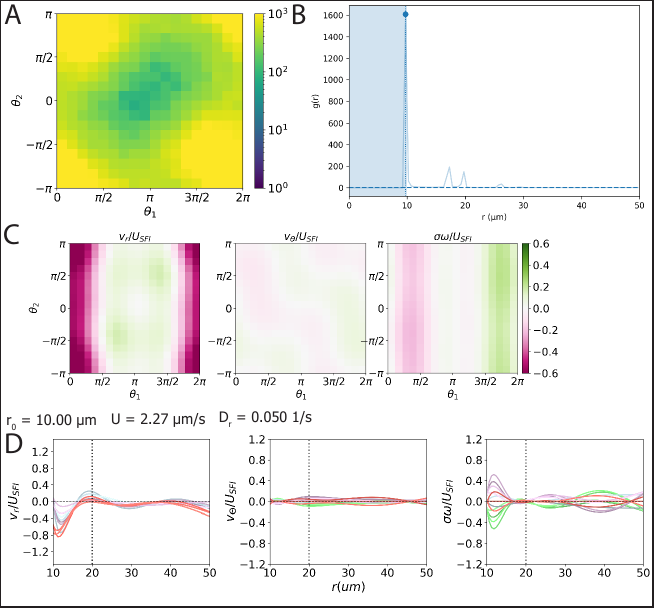}
  \caption{Inference results for a simulation without exponential repulsion, only a soft repulsive soft disk contact force, which acts when particles start overlapping, \textit{i.e.}~when $r_0 \leq 10$ \si{\micro\meter}. (A) Pair correlation function for all relative orientations at $10$ \si{\micro\meter} (B) Pair correlation function over the cell length (C) Inferred normalized pair interactions inter-particle distance $r_{0}=10$ \si{\micro \meter} as a function of $\theta_1$ and $\theta_2$. (D) Normalized $v_{r}$, $v_{\theta}$ and $\omega$ as a function of the radial distance for different angle combinations. The vertical line marks $r_0=2\sigma$. The color code is the same as used in Figure S7.}
  \label{fgr:fig9SI}
\end{figure}

\clearpage
\subsection{Simulation without phoretic pull}
\begin{figure}[h]
\centering
  \includegraphics{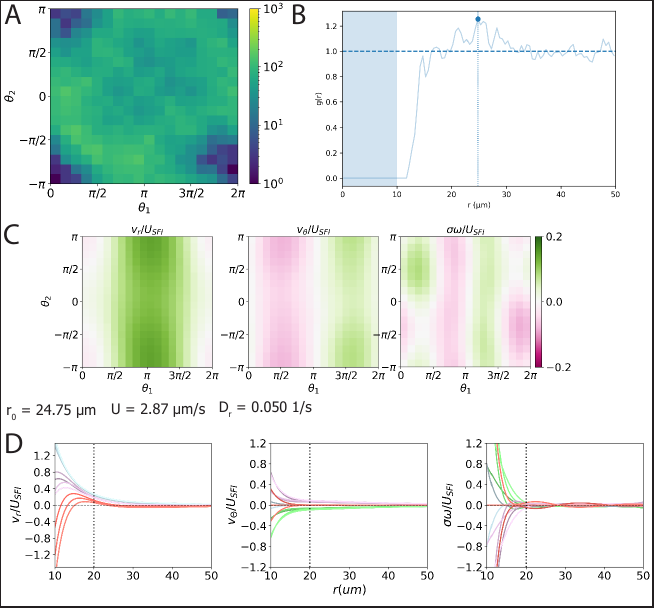}
  \caption{Inference results for a simulation without phoretic pull. (A) Pair correlation function for all relative orientations at $24.75$ \si{\micro\meter} (B) Pair correlation function over the cell length (C) Inferred normalized pair interactions inter-particle distance $r_{0}=24.75$ \si{\micro \meter} as a function of $\theta_1$ and $\theta_2$. (D) Normalized $v_{r}$, $v_{\theta}$ and $\omega$ as a function of the radial distance for different angle combinations. The vertical line marks $r_0=2\sigma$. The color code is the same as used in Figure S7.}
  \label{fgr:fig_no_phoretic_SI}
\end{figure}
\clearpage
\subsection{Simulation with stronger $J$}
\begin{figure}[h]
\centering
  \includegraphics{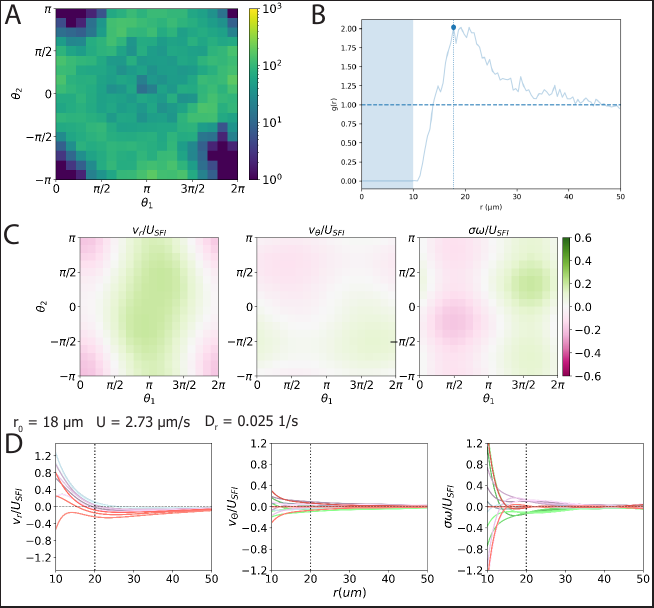}
  \caption{Inference results for a simulation with a higher phoretic pull strength $J$. $J$ has a value of 5.0. (A) Pair correlation function for all relative orientations at $18$ \si{\micro\meter} (B) Pair correlation function over the cell length (C) Inferred normalized pair interactions inter-particle distance $r_{0}=18$ \si{\micro \meter} as a function of $\theta_1$ and $\theta_2$. (D) Normalized $v_{r}$, $v_{\theta}$ and $\omega$ as a function of the radial distance for different angle combinations. The vertical line marks $r_0=2\sigma$. The color code is the same as used in Figure S7.}
  \label{fgr:fig_higher_J_SI}
\end{figure}
\clearpage
\subsection{Simulation with centered catalytic kernel}
\begin{figure}[h]
\centering
  \includegraphics{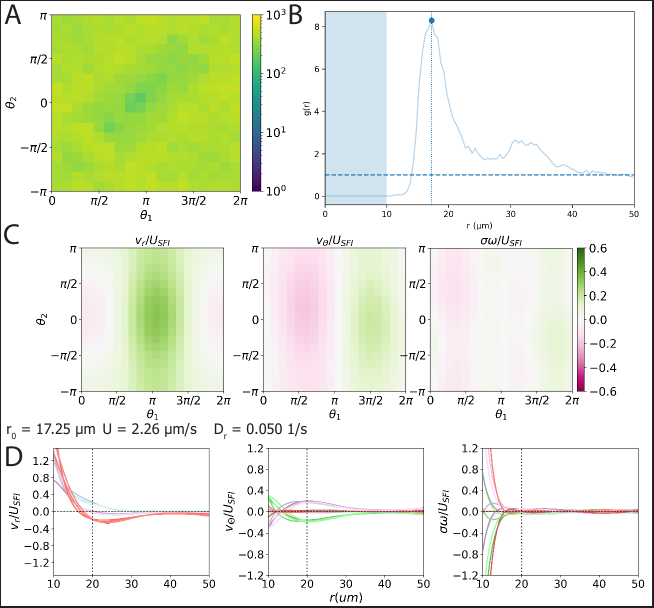}
  \caption{Inference results for a simulation with a centered catalytic kernel. (A) Pair correlation function for all relative orientations at $17.25$ \si{\micro\meter} (B) Pair correlation function over the cell length (C) Inferred normalized pair interactions inter-particle distance $r_{0}=17.25$ \si{\micro \meter} as a function of $\theta_1$ and $\theta_2$. (D) Normalized $v_{r}$, $v_{\theta}$ and $\omega$ as a function of the radial distance for different angle combinations. The vertical line marks $r_0=2\sigma$. The color code is the same as used in Figure S7.}
  \label{fgr:fig_no_shift_SI}
\end{figure}

\clearpage